\documentstyle[aps,twocolumn,titlepage,epsf,psfig]{revtex}
\begin{document}
\newcommand {\mb  } {\mbox    }
\newcommand {\al  } {\alpha   }
\newcommand {\be  } {\beta    }
\newcommand {\s   } {\sigma   }
\newcommand {\la  } {\lambda  }
\newcommand {\de  } {\delta   }		
\newcommand {\De  } {\Delta   }		
\newcommand {\eps } {\varepsilon       }
\newcommand {\laik} {{\la_{i_k}}       }
\newcommand {\vlal} {{\vec{\la_l}}     }
\newcommand {\erfc} {\mb{${\rm erfc }$}}
\newcommand {\ha  } {\mb{${1\ov2}$}   }
\newcommand {\ola } {{\overline{\la }}}
\newcommand {\e   } {\!+\!    }
\newcommand {\m   } {\!-\!    }
\newcommand {\lgl } {\langle  }		
\newcommand {\rgl } {\rangle  }		
\newcommand {\lc  } {\left\{  }
\newcommand {\lp  } {\left .  }
\newcommand {\rp  } {\right . }
\newcommand {\lh  } {\left(   }
\newcommand {\rh  } {\right)  }
\newcommand {\lv  } {\left[   }
\newcommand {\rv  } {\right]  }
\newcommand {\llgl} {\left\langle    }
\newcommand {\rrgl} {\right\rangle   }
\newcommand {\beq } {\begin{equation}}
\newcommand {\eeq } {\end{equation}  }
\newcommand {\bea } {\begin{eqnarray}}
\newcommand {\eea } {\end{eqnarray}  }
\newcommand {\bay } {\begin{array}   }
\newcommand {\eay } {\end{array}     }
\newcommand {\noi } {\noindent   }
\newcommand {\ov  } {\over       }
\newcommand {\pa  } {\partial    }
\newcommand {\hsp } {\hspace     }
\newcommand {\np  } {\newpage    }
\newcommand {\nl  } {\newline    }
\newcommand {\nn  } {\nonumber   }
\newcommand {\lb  } {\label      }
\newcommand {\fa  } {\forall     }
\newcommand {\ev  } {\equiv      }
\newcommand {\n   } {\hat{n}     }
\newcommand {\q   } {\hat{q}     }
\renewcommand {\vr  } {{\vec{r}}   }
\newcommand {\hcB } {\hat{\cal B}}
\newcommand {\hcN } {\hat{\cal N}}
\newcommand {\hcP } {\hat{\cal P}}
\newcommand {\cD  } {{\cal D}}
\newcommand {\cF  } {{\cal F}}	     
\newcommand {\cG  } {{\cal G}}
\newcommand {\cH  } {{\cal H}}	     
\newcommand {\cK  } {{\cal K}}
\newcommand {\cN  } {{\cal N}}	     
\newcommand {\cO  } {{\cal O}}	     
\newcommand {\cS  } {{\cal S}}	     
\newcommand {\cZ  } {{\cal Z}}	     
\newcommand {\hB  } {\hat{B} }
\newcommand {\hM  } {\hat{M} }
\newcommand {\hS  } {\hat{S} }
\def\Tr{\mathop{\mb{\rm Tr}}}
\twocolumn[
\hsize\textwidth\columnwidth\hsize\csname @twocolumnfalse\endcsname
\title{Solvable Models of Random Hetero-Polymers at Finite Density: I.
Statics}

\author{Jort van Mourik \\
 Department of Mathematics,
 King's College London, Strand, London WC2R 2LS, UK,\ and \\
 Istituto Nazionale di Fisica della Materia,
 Via Beirut 2-4, 34014 Trieste, Italy.                    \\
 email: jvmourik@mth.kcl.ac.uk
}

\maketitle

\begin{abstract}
\noi We introduce $\infty$-dimensional versions of three common models of 
random hetero-polymers, in which both the polymer density and the density 
of the polymer-solvent mixture are finite. These solvable models give 
valuable insight into the problems related to the (quenched) average over 
the randomness in statistical mechanical models of proteins, without having
to deal with the hard geometrical constraints occurring in finite 
dimensional models. Our exact solution, which is specific to the 
$\infty$-dimensional case, is compared to the results obtained by a 
saddle-point analysis and by the grand ensemble approach, both of which can
also be applied to models of finite dimension. We find, somewhat 
surprisingly, that the saddle-point analysis can lead to qualitatively 
incorrect results.
\end{abstract}
\vskip 3mm

June 15, 1999 
\vskip 1cm
]

\section{Introduction \lb{int}}

The study of random hetero-polymers is regarded as a natural first step
towards a better understanding of the physics of proteins \cite{Cr}. The
latter are believed to be {\em special} realisations of random
hetero-polymers which have been selected by nature; before tackling these
a-typical (selected) cases, one first investigates the properties of
typical hetero-polymers. Several models of such systems have been proposed
and studied (for an overview see e.g. \cite{GOT,PGT}). Most involve a
coarse-grained description of the polymer chain, where each monomer (which
may consist of many atoms) is represented as a single bead, and {\em
effective} two-body interactions between monomers and/or solvent molecules.
Since proteins are hetero-polymers of fixed composition, one is mostly
interested in the case of quenched disorder. Yet, studying the case of
annealed disorder can also be relevant,  since it may describe a situation
in which the monomers are able to change their chemical properties (e.g. by
undergoing chemical reactions or by exchanging charge), and secondly (at a
technical level) since one can approximate the quenched result by a series
of annealed averages following the scheme of Morita \cite{Mo}.

The principal objective of this paper is to clarify the problems that occur
when following a saddle point (SP) approach in dealing with the (quenched)
randomness in those coarse grained models of random hetero-polymers which
take into account the incompressibility of the solvent-monomer system.
Although the behaviour of random hetero-polymers in finite dimensions will
obviously differ from those in high dimensions (due to the crucial role in
the former of the chain constraint), we find that the study of infinite
dimensional models provides valuable insight into the appropriateness of
some of the approximations and assumptions that are usually being made, as
it disentangles the problems induced by the disorder from those generated
by the  configurational entropy. It is beyond the scope of this work to
discuss the numerous variational approaches that have been introduced,
often in combination with a saddle-point treatment. Neither do we want to
go into details about possible replica symmetry breaking (RSB) that may
occur in some models. Although the dominance of the entropy in our present
models does not permit real phase transitions at finite temperatures, we do
observe phase coexistence and, especially in the case of annealed disorder,
(re-entrant) cross-over between swollen and compact phases, which may
indicate real (re-entrant) phase transitions in the corresponding finite
dimensional models.

This paper is organised as follows. In section II we introduce our
models, and illustrate connections with related models such as those
describing matching problems and poly-dispersity. In sections III
and IV we derive the (exact) solution for the case of discrete
disorder, and for arbitrary disorder distributions (in terms of
mesoscopic variables), respectively. In section V we study the
same models with a (na\"{\i}eve) mean field approach, and point out
the latter's weaknesses. In section VI we follow the grand ensemble
approach of \cite{Mo}, where the quenched average is approximated by a
series of annealed averages. Although hard to implement for real quenched
disorder, this scheme may quite accurately describe polymers with
so-called permuted disorder, where the monomers are allowed to
migrate along the chain. Finally, in section VII, we discuss our results
 and present an outlook on future work.

\section{Model Definitions \lb{mod}}

\noi
We consider large chains of random hetero-polymers. The position of each
monomer $i=1,..,N$ along a chain is given by a vector $\vr_i$. A function
$g(\{\vr\})$ will incorporate all the constraints on the possible
configurations of the chain, such as self-avoidance, bond length, lattice
constraints etc., such that the free energy of the polymer can be written
as

\beq
\cF=-{1\ov\be}\log\lv\lh\prod_{i=1}^N\Tr_{\vr_i})\rh g(\{\vr\})
\exp(-\be \cH(\{\vr\}))\rv .
\label{eq:free_energy}
\eeq

\noi In the case of short range 2-body interactions, the interaction
Hamiltonian will be of the form

\beq
\cH(\{\vr\})=\sum_{i<j=1}^N \lambda_{ij} \de_{\vr_i,\vr_j}\ .
\label{eq:hamiltonian}
\eeq

\noi If the interaction energies $\lambda_{ij}\in\Re$ are drawn
independently for each pair $(i,j)$ with $i<j$, we have a model with
bond-disorder. Alternatively we could choose for each monomer to carry a
set of labels $\ola_i\in\Re$, specifying its electro-chemical properties,
and define the interaction energies as fixed functions of the labels of the
monomers involved, i.e. $\lambda_{ij}=\Lambda(\ola_i,\ola_j)$, which would
lead to a model with site disorder.

Since (\ref{eq:free_energy}) cannot be calculated for all possible
disorder realisations, and since the free energy is a self-averaging
quantity (i.e. the average value is the typical value), one is interested
in its average over the disorder, the so-called quenched free energy
$\cF_q=\lgl\cF\rgl_{\{\lambda_{ij}\}}$\ .
In order to investigate the influence of disorder
in the interactions between the different monomers, without having to
make additional approximations to deal with the configurational entropy,
we consider the following simplified $\infty$-dimensional model:
\begin{itemize}
\item
There are $N$ monomers labeled $i=1,..,N$\ , and $R$ sites labeled
$r=1,..,R$.
\item
Every site is a neighboring site to every other site, and thus the chain
constraint of the polymer is always trivially satisfied.
\item
Each site can contain up to $n_c$ monomers. This accounts for the
incompressibility of the polymer-solvent system, and restricts the global
density according to $N/R\leq n_c$, with equality corresponding to the
situation where all sites are completely filled with monomers.
\item
Those positions at a given site which are not occupied by monomers, are
occupied by solvent molecules.
\item
Only those particles interact which find themselves in the same site.
\end{itemize}

\begin{figure}[t]
\setlength{\unitlength}{0.55mm}
\begin{picture}(0,70)
\put(-10, 5){\epsfxsize=160\unitlength\epsfbox{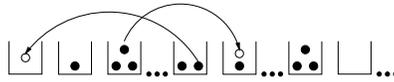}}
\end{picture}
\vspace*{3mm}
\caption{
Illustration of the $\infty$-dimensional model, for $n_c=3$,
with some of the allowed moves indicated with arrows.
The upper configuration is one where the sites are not yet
ordered; the lower
configuration is obtained upon ordering the sites according to
monomer occupation numbers.}
\lb{fg1}
\end{figure}

\noi The problem of calculating the free energy has now been reduced to a
relatively simple combinatorial one. The number of monomers at site $r$ is
denoted by $n_r$. Hence, in terms of the notation of
(\ref{eq:free_energy}), $g(\{r\})=1$ for all configurations with
$n_r\leq n_c$ for all $r$, and $0$ otherwise:

\beq
g(\{r\})=\prod^R_{r=1}\lh1-\sum_{k>n_c}\de_{k,n_r(\{r\})}\rh\ .
\eeq

\noi Note that we must always satisfy the following relations

\beq
n_r\ev\sum_i\de_{r,r_i}    \ ,\quad
\sum_r n_r=N               \ ,\quad
N\leq\sum_in_{r_i}\leq n_cN\ .
\lb{nr}
\eeq

\noi The problem is symmetric under arbitrary permutations of the sites.
We now order these such that the first $C_{n_c}N$ sites contain $n_c$
monomers, the next $C_{n_c\m1}N$ contain $n_c\m1$ monomers, etc. (see
FIG. \ref{fg1}). The last $C_0 N$ sites contain no monomers. We define
the total fraction of monomers that find themselves at a site which is
occupied by $l$ monomers as $f_l\ev lC_l,\ l=1,..,n_c$. Thus

\beq
\sum_{l=0}^{n_c}lC_l=\sum_{l=1}^{n_c}f_l=1\ ,\quad
\sum_{l=0}^{n_c}C_l=\al        \ .
\lb{cf}
\eeq

\noi A monomer at a site occupied with $n_c$ monomers is said to be in the
(maximally) compact state, a monomer at a site where there are no other
monomers is said to be in a swollen state, and monomers which find
themselves at sites  with an intermediate number of monomers are in a
semi-compact state. The macroscopic state of the polymer is characterised
by the numbers $\{C_l,l=0,..,n_c\}$ (which are proportional to the
fractions of sites with a given monomer occupation number). The overall
compactness of the polymer can be measured by a single parameter $D$,
with $D=0$ describing a maximally swollen configuration and
$D=1$ describing a maximally compact one:

\beq
D\ev{(1-\alpha+C_0)\ov(1\m1/nc)}
\eeq

\noi By varying $\al$, the total monomer density and the fraction of
monomers in (semi-)compact states at high temperatures can be made
arbitrarily small. The state where all monomers are found at the same
finite density $n_c$ is the high dimensional version of the compact
globular phase for random hetero-polymers, which is often considered in the
literature. Note that it is not appropriate to view the present models as
coarse-grained high dimensional limit of lattice models, although the short
range nature of the interactions is mimicked by representing small
elementary volumes by sites and taking only (delta) interactions within
this volume, ignoring the interactions between these elementary volumes.
For a true high dimensional limit, the number of sites in an elementary
volume would increase exponentially with the dimension, while here it is
kept finite. Hence, it is rather a geometry-free version of the
finite dimensional continuum models that have been considered.

We can now calculate typical quantities for boxes containing $l$ monomers,
provided we add the entropic contribution $S_{\rm ord}$ to the free energy
which reflects our ordering of sites, given by the logarithm of the
associated combinatorial factor:

\beq
\log\lh{(R)!\ov\prod_{l=0}^{n_c}(C_l N)!}\rh\simeq
N[\ \al\log(\al)-\sum_{l=0}^{n_c}C_l\log(C_l)\ ]\ ,
\lb{ent}
\eeq

\noi (using Stirling's formula). As a consequence of the fractions
$\{C_l\}$ being a finite number of intensive macroscopic observables, we
will be able to solve our models by minimizing the free energy with respect
to the $\{C_l\}$ (or, equivalently the $\{f_l\}$). Alternatively, we can
separate the total entropy $S$ into two contributions: the entropy
$S_{\rm g}$ of grouping the $N$ monomers,

\bea
&&\!\!\log\lh{N!\ov\prod_{l=1}^{n_c}(C_lN)!\ (l!\ )^{C_lN}}\rh= \\
&&N(1\m\sum_{l=1}^{n_c}C_l)\log(N)\m N(1\m\sum_{l=1}^{n_c}C_l)\m
N\sum_{l=1}^{n_c}C_l\log(l!\ C_l))\ .\nn
\eea

\noi and, secondly, the residual entropy $S_{\rm p}$ describing the
remaining freedom in positioning the monomers at the $\al N$ sites:

\bea
&&\!\!\log\lh{(\al N)!\ov(C_0 N)!}\rh= \\
&&(\al\m C_0)N\log(N)\e N[\al\log(\al)\m C_0\log(C_0)\m(\al\m C_0)]\ .
\nn
\eea

\noi Hence, the total entropy $S$ is given by

\beq
S_{\rm g}\e S_{\rm p}=
N\log(N)\m N\e N[\al\log(\al)\m\sum_{l=0}^{n_c}C_l\log(l!\ C_l)]
\eeq

\noi Since $S_{\rm g}$ contains a term proportional to $N\log(N)$, the
entropy will always dominate the free energy, except at properly rescaled
(very low) temperatures. This was already recognized by Vannimenus et al.
\cite{VM} for so-called matching problems. In what follows we will
systematically omit all contributions to the entropy which are independent
of the configuration; the entropy may consequently become negative at low
temperatures, and non-monotonicity of the free energy need no longer
indicate a phase transition in the system.

Our model can be interpreted as a hetero-polymer on a simplex in $\al N\m1$
dimensions. It differs, though, from the infinite dimensional random
Potts model, firstly through the fact that the (quenched) randomness is
not in the background i.e. in the sites or in interactions coupling sites
(in fact there are no couplings between different sites), but in the
(mobile) monomers, and secondly since only a finite number of
monomers are allowed to be at the same site.
For the latter reason, and because the energy per site depends on the
(number of) particles present at that site, it also differs from the {\em
back-gammon} model of Ritort \cite{Ri}. However, it shares the property
that the entropy is not extensive.
Our model can be seen as similar to those describing matching problems
(with here monomers as the objects to be `matched').
Note that in standard matching problems either the density is fixed
(typically at $n_c=2$) \cite{VM,Or,MP}, or properly scaled chemical
potentials are introduced \cite{N}, both with the objective to keep the
non-extensive (permutational) part of the entropy constant.
In contrast, in the present model the latter is achieved by limiting the
number of available sites to $\al N$. Although the interesting behaviour
of matching problems (such as freezing) only occurs at properly rescaled
(extremely low) temperatures, and a lower bound on the interaction
distribution is needed to have a finite ground-state energy, we have not
investigated this temperature regime in the present paper, but in analogy
with those papers on random hetero-polymer models whose assumptions and
methods we wish to investigate, we have considered continuous disorder
distributions of a Gaussian form.
Finally, one can also view our model as one describing the coexistence of
a large number (or infinite number, in the case of continuous disorder
distributions) of monomer species, i.e. as a poly-disperse mixture (see
e.g. \cite{poly} and references therein).

What remains to fully specify our models is to make explicit choices for
the pair-interactions between monomers.
In the present paper we consider
the three most commonly used interaction types:
\begin{itemize}
\item
In the {\em random hydrophobic/hydrophilic model} (RHM)
\cite{Ob,GLO,TAB,TvMM} a label $\la_i$ is assigned to each monomer (with
$\la_i>0$ for hydrophilic and $\la_i<0$ for hydrophobic monomers), and

\bea
&&\lambda_{ij}=\lambda_i+\lambda_j~~(\la_i~~{\rm i.i.d.r.v.}):
\lb{HH} \\
&&\cH(\{\vr\})=\frac{1}{2}\sum_{i\neq j}^N [\lambda_{i}+\lambda_j]
\de_{r_i,r_j}=\sum_i\la_i\ n_{r_i}+{\rm const}\nn
\eea

\noi Note that this is only the configuration dependent term of the full
interaction energy. Each contact of a monomer $i$ with a solvent molecule
contributes $-\la_i$ to the energy (hence the name hydrophilicity).
The full Hamiltonian is thus given by
$\cH=-\sum_i\sum_s\la_i\de_{r_i,r_s}=$$-\sum_i\la_i(n_c-n_{r_i})$\, with
the index $s$ running over the solvent molecules.
\item
The {\em random charge model} (RCM) \cite{GO_C,Sf,GK} (or
`random sequence' model) describes inter-monomer Coulomb-like
interactions. Again each monomer carries a label $\la_i$ (its `charge').
The contribution to the energy is proportional to the product of
the charges of monomers in contact, such that equally signed charges either
repel ($\epsilon=1$) or attract ($\epsilon=-1$) each other:

\bea
&&\lambda_{ij}=\epsilon\lambda_i\lambda_j~~(\la_i~~{\rm i.i.d.r.v.}):
\lb{HC}\\
&&\cH(\{\vr\})={\eps\ov2}\sum_{i\neq j}^N\la_i\la_j\de_{r_i,r_j}\nn
={\eps\ov2}\sum_r\la_r^2+{\rm const}
\eea

\noi where $\la_r\ev\sum_i\la_i\de_{r,r_i}$ is the total charge in box
$r$, and $\eps=\pm1$. Solvent molecules are assumed to be neutral.
\item
In the {\em random interaction model} (RIM) \cite{BW,GO_I,SG}
(or `random bond' model) the $\frac{1}{2}N(N-1)$
interactions $\lambda_{ij}$ between the
pairs of monomers are themselves taken to be independent identically
distributed random variables (i.i.d.r.v.), drawn from a probability
distribution $P(\la)$, which may either be discrete or continuous:

\bea
&&\lambda_{ij}~~{\rm i.i.d.r.v.}~~(\la_{ji}\ev\la_{ij}): 
\lb{HI}\\
&&\cH(\{\vr\})={1\ov2}\sum_{i\neq j}^N\la_{ij}\de_{r_i,r_j}\ .\nn
\eea

\end{itemize}

\section{Discrete Disorder: Exact Solutions \lb{DD}}

\noi In this section we deal with the case where the interactions take
discrete values, i.e. where the $\lambda_i$ (for RHM and RCM) or the
$\lambda_{ij}$ (for RIM) have probability distributions of the form

\beq
P(\la)=\sum_{\s=1}^{N_\s}p_\s\de(\la-\la_\s)\ ,
\lb{Pd}
\eeq

\noi with finite $N_\s$.
We define `local states' as the monomer configurations that can occur at a
given site (for site-disordered models these local states are fully
characterised by the number of monomers of a given type present).
Note that for finite $n_c$ and finite $N_\s$ the number $N_{\rm ls}$
of possible local states is also finite. Furthermore, due to the nature of
the Hamiltonian (\ref{eq:hamiltonian}), the total energy is just the sum
of the local energies in at the $\alpha N$ sites. We define the
concentration $c_{s_l}$ of a local state $s_l$ as the number of times it
occurs among the $\alpha N$ sites ($s_l=1,..,N_{\rm ls})$ divided by $N$,
and the number of monomers occupying a site in local state $s_l$ by
$n_{s_l}$. We then always have to satisfy the normalisation requirements
\beq
\sum_{s_l}c_{s_l}=\al\ ,\hsp{1cm}\sum_{s_l}n_{s_l}\ c_{s_l}=1\ .
\lb{nor}
\eeq

\subsection{RHM \& RCM}

\noi For the (site-disordered) RHM and RCM models, one can define the
local state at a given site  by a set of numbers $\{m^\s\}$
$(\s=1,..,N_\s)$, where each  $m^\s$ specifies the number of monomers
present of type $\s$. In this way we automatically take care of the
invariance under permutations of monomers at the same site. The
concentration $c_{\{m^\s\}}$ of a local state $\{m^\s\}$ is the number of
times it occurs among the sites, divided by $N$. In addition to (\ref{nor})
we now have $N_\s\m1$ extra independent normalisation conditions,
reflecting conservation of monomer species:

\beq
\sum_{\{m^\s\}}m^{\s'}\ c_{\{m^\s\}}=p_{\s'}\ ,\hsp{1cm}\fa\s'=2,..,N_\s \ .
\lb{norP}
\eeq

\noi The total  number of possible local states is given by

\beq
N_{\rm ls}=\sum_{l=0}^{n_c}\lh\!\!\bay{c} N_\s\e l\m1\\ l \eay\!\!\rh=
\lh\!\!\bay{c}N_\s\e n_c\\ n_c\e1\eay\!\!\rh=
{(N_\s\e n_c)!\ov(n_c\e1)!(N_\s\m1)!}\ .
\lb{Nu}
\eeq

\noi It is now straightforward to write down the state-dependent part
of the free energy, for large $N$:

\bea
\cF_{\rm sd}&=&(E-TS)/N=\sum_{\{m^\s\}}c_{\{m^\s\}}\lv
E(\{m^\s\})+\hsp{-30cm}{1\ov2}\rp\lb{Fd}\\
&&~~~~\lp+{1\ov\be}\log\lh(\prod_\s m^\s!)\ c_{\{m^\s\}}\rh\rv
~~~~~(N\to\infty),\nn
\eea

\noi and the models are exactly solved at all temperatures.
This expression is to be minimized with respect to the concentrations
$c_{\{m^\s\}}$ (of which there is only a finite number, even for
$N\to\infty$, and which play the role of order parameters), under the
constraints (\ref{nor},\ref{norP}). This procedure is in principle
straightforward, however, since the number of order parameters (i.e. the
number of $c_{\{m^\s\}}$), as given by (\ref{Nu}), grows considerably with
the number of monomer species $N_\sigma$ and with $n_c$, we have restricted
ourselves to the case $n_c=3$ and $n_\s=2$ (bimodal disorder:
$\la_i=\pm1$). This case turns out to already exhibit all the essential
features of our models. The results of determining the order parameters
$\{c_{\{m^\s\}}\}$ by minimization of (\ref{Fd}), are compared with
numerical simulations in FIG. \ref{fg2} and FIG. \ref{fg3}, for the RHM and
the RCM respectively.

\begin{figure}[t]
\vspace*{10mm}
\setlength{\unitlength}{0.6mm}
\begin{picture}(0,90)
\put( -8,10){\epsfxsize=140\unitlength\epsfbox{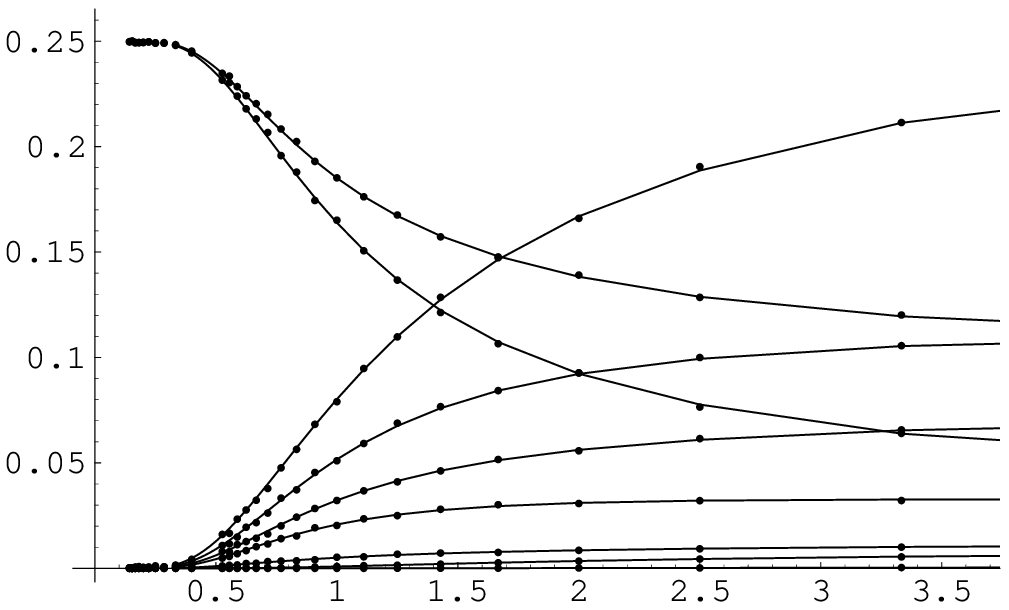}}
\put( 45,75){\makebox(3,1){\large $c_{\{0,1\}}$}}
\put( 30,60){\makebox(4,1){\large $c_{\{3,0\}}$}}
\put( 67,10){\makebox(0,-5){\large $T$}}
\end{picture}
\vspace*{5mm}
\caption{
Order parameters $c_{\{n_-,n_+\}}$  as functions of temperature, for the
Random Hydrophobic/Hydrophilic Model (RHM),
with $n_c=3$, $\alpha=1$, $\la_i=\pm1$ and $p(\lambda_i=1)=\frac{1}{4}$. Full lines: theory.
Dots: numerical simulations ($N=3000$).}
\lb{fg2}
\end{figure}

\noi We clearly observe the tendencies one expects. In the
RHM model the low temperature local configurations are ones where either a
site is fully filled with hydrophobic monomers (which thereby avoid contact
with solvent molecules),  

\begin{figure}[t]
\vspace*{10mm}
\setlength{\unitlength}{0.6mm}
\begin{picture}(0,170)
\put( -2,100){\epsfxsize=140\unitlength\epsfbox{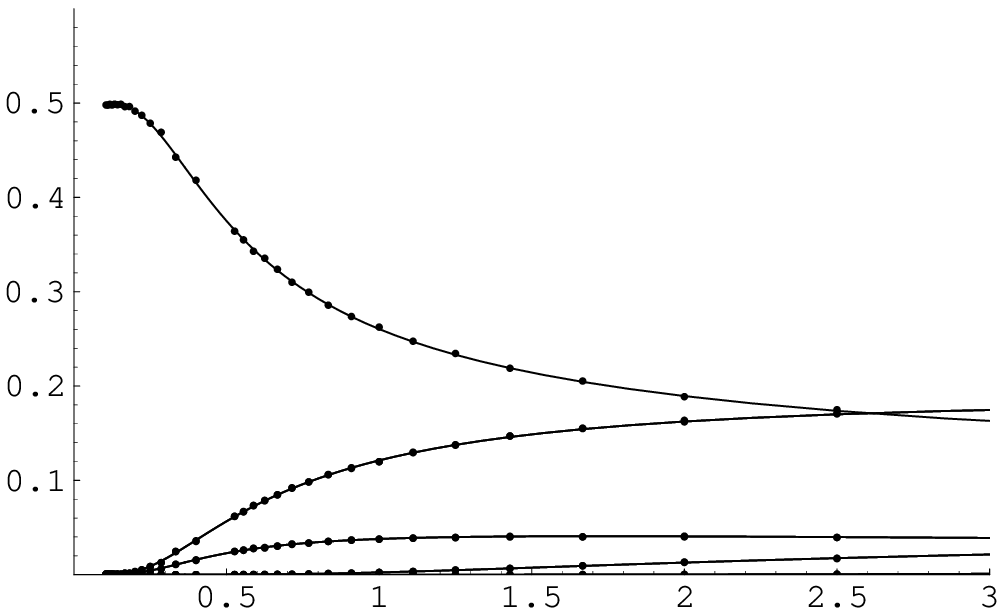}}
\put( -8,  7){\epsfxsize=140\unitlength\epsfbox{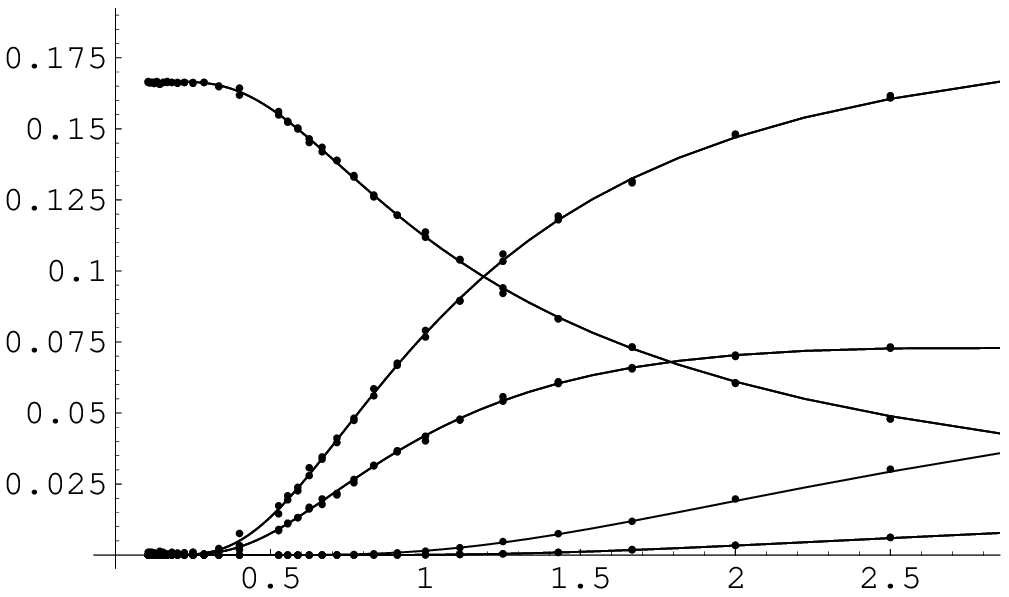}}
\put( 52, 76){\makebox(3, 1){\large $c_{\{3,0\}}=c_{\{0,3\}}$}}
\put( 47,150){\makebox(3, 1){\large $c_{\{1,1\}}$}}
\put( 67,  7){\makebox(4,-4){\large $T$}}
\put( 67, 98){\makebox(4,-4){\large $T$}}
\end{picture}
\vspace*{5mm}
\caption{
Order parameters $c_{\{n_-,n_+\}}$  as functions of temperature, for the
Random Charge Model (RCM), with $n_c=3$, $\alpha=1$, $\la_i=\pm1$ and
$p(\lambda_i=\pm 1)=\frac{1}{2}$. Upper graph: $\epsilon =1$ (equal
charges repel), lower graph: $\epsilon=-1$ (equal charges attract).
Full lines: theory. Dots: numerical simulations ($N=3000$).}
\lb{fg3}
\end{figure}

\noi or contains a single hydrophilic monomer (which then maximises the 
contacts with solvent). For the RCM we see that at low temperatures for 
$\epsilon=1$ (equal charges repel) the system favors states where opposite 
charges pair, whereas for $\epsilon=-1$ (equal charges attract) the 
preferred states are the ones where equally charged monomers maximally 
group together. In all cases the agreement between theory and experiment is
excellent.

\subsection{RIM}

\noi In order to solve the Random Interaction Model (RIM) we have to follow
a slightly different strategy.
Firstly, because the local state at a site $r$ is now defined by the
number $n_r$ of monomers situated at $r$, in combination with the
$n_r\times n_r$ traceless interaction matrix $M_r$ of the interaction
energies between pairs of these monomers. The number of independent matrix
elements of $M_r$ is $n_r(n_r\m1)/2$, each can take all of the $N_\s$
different values, giving rise to $N_\s^{n_r(n_r\m1)/2}$ different possible
local matrices. Since we can freely permute the monomers at a given site,
many of these matrices give rise to the same local state, and it is
sufficient to select a prototype for each of the non-equivalent classes.
Given a local interaction matrix $M$, we can divide the monomers into
$N_t$ classes of size $m^t,\quad t=1,..,N_t$, such that $M$ is invariant
under all permutations within the same class, and not so for any
permutation of two monomers from different classes.
Secondly, because only a negligible fraction of the labels $\la_{ij}$ are
actually visible in the interactions, we cannot impose the equivalent of
the condition in (\ref{norP}), but instead we use the fact that the
ensemble probability that a label $\la$ occurs, is given by $P(\la)$. Note
that this equivalent to assuming that the annealed approximation of the
quenched average is valid; in appendix {\bf B} we will prove the validity
of this assumption at finite temperatures.
The state-dependent part of the free energy is now asymptotically given by

\bea
\cF_{\rm sd}&=&(E-TS)/N=\sum_{(n,M)}c_{(n,M)}\lv E(n,M)+
\hsp{-30cm}{1\ov2}\rp \lb{Fdi}\\
&&~~~\lp+{1\ov\be}\log\lh{\prod_tm^t!\ c_{(n,M)}\ov\prod_\s(p_\s)^{m^\s_M}}
\rh\rv~~~~~~(N\to\infty),\nn
\eea

\noi where $m^\s_M$ is the number of times the interaction $\s$ occurs one
entry above the diagonal in the matrix $M$, and the model is exactly solved
at finite temperatures.

\begin{figure}[t]
\setlength{\unitlength}{0.6mm}
\vspace*{5mm}
\begin{picture}(0,85)
\put(-8, 5){\epsfxsize=140\unitlength\epsfbox{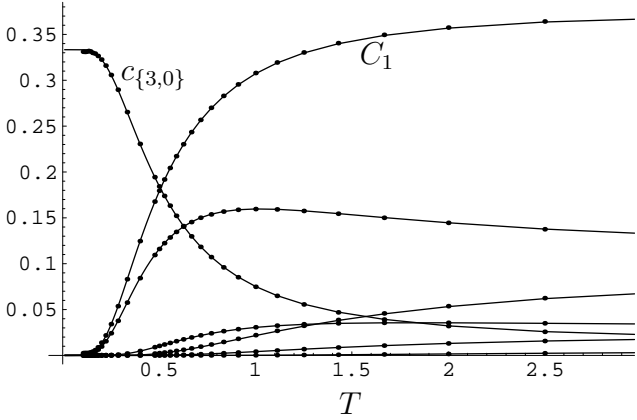}}
\put(24,73){\makebox(3,1){\large $c_{\{3,0\}}$}}
\put(73,78){\makebox(4,1){\large $C_1$}}
\put(67, 1){\makebox(4,1){\large $T$}}
\end{picture}
\vspace*{5mm}
\caption{
Order parameters $c_{\{n_-,n_+\}}$ and $C_1$ as functions of temperature,
for the Random Interaction Model (RIM), with $n_c=3$, $\alpha=1$,
$\la_{ij}=\pm1$ and $p(\lambda_{ij}=\pm 1)=\frac{1}{2}$. Full lines:
theory. Dots: numerical simulations ($N=3000$).}
\lb{fg4}
\end{figure}

\noi The order parameters $c_{(n,M)}$ follow from minimisation of
(\ref{Fdi}), subject to the constraints (\ref{nor}). For general $n_c$ and
$N_\s$ it is very cumbersome to determine the number of local states
$N_{\rm ls}$. Therefore, we have again restricted ourselves to the simple
case $n_c=3$ (for which the $(n,M)$ are fully determined by the numbers
$\{m^\s_M\}$, whence $c_{\{m^\s_M\}}$), $N_\s=2$ with $N_{\rm ls}=8$; the
result of comparing the predictions of the theory with numerical
simulations is shown in FIG. \ref{fg4}.

\section{Continuous Disorder: Exact Solutions}

\noi In the general case of arbitrary and possibly continuous disorder
distributions, we introduce densities $c(l,\vlal)$ of boxes with a
specified number $l$ of monomers and given labels
$\vlal$ (representing local attributes such as hydrophilicities, charges or
sub-matrices of interactions, depending on the model at hand). Since the
For $N\to\infty$ we assume the fraction of sites with $l$ monomers and
$\vlal\in[\vlal,\vlal+\De\vlal]$ to be given by $c(l,\vlal)\De\vlal$,
where $c(l,\vlal)$ depends continuously on $\vlal$ and is self-averaging.
For these assumptions to be true the dimension of $\vlal$ must be small
compared to $N$. The $c(l,\vlal)$ are so-called mesoscopic variables. Note
that our approach is similar to that of the sublattice magnetisations as
introduced for spin models in \cite{HGHK}, and applies for as long
as the number of possible labels $\vlal$ is sufficiently small to guarantee
that the number of members in each class of momoners with prescribed labels
is infinite, and that the fluctuations in the $c(l,\vlal)$ will not
contribute to the free energy.

\subsection{RHM \& RCM}

\noi For the RHM and the RCM, the solution involves only minor
adaptations/generalisation of the one obtained for discrete disorder. The
asymptotic state-dependent contribution to the free energy can formally be
written as a functional over the above densities, where
$\vlal\ev\{\la_{i_k},\ k=1,..,l\}$ are the labels of the $l$ monomers
present at a given site:

\bea
&&{\cal F}_{\rm sd}=
\sum_{l=0}^{n_c}\int d\vlal\ c(l,\vlal)\lv E(l,\vlal)+
{1\ov\be}\log(l!\ c(l,\vlal))\rv\ ,\nn\\
&&\hsp{6cm}(N\to\infty)
\lb{Free}
\eea

\noi where $d\vlal\!\ev\!(\prod_{k=1}^ld\laik)$ and where $E(l,\vlal)$ is
the contribution to the system's total energy from a site with $l$ monomers
characterised by label vector $\vlal$. The term $-c(l,\vlal)
\log(l!)$ in the entropy reflects the fact that the $c(l,\vlal)$ are
invariant under permutations of the $\laik,\ \ k=1,..,l$. In the case of
there being a non-vanishing probability for the monomers to be identical,
i.e. when the probability distribution of the labels contains delta peaks
(as for discrete disorder) we have to add to the entropy the term
$-c(l,\vlal)\log(\prod_\la n_\la!)$, where $\la$ runs over the different
values which the components of $\vlal$ take, and where
$n_\la\!\ev\!\sum_{k=1}^l\de_{\laik,\la}$\ .
The equilibrium values of the concentrations $c(l,\vlal)$ are to be
determined via minimization of (\ref{Free}), subject to the constraints

\bea
1&=&\sum_{l=1}^{n_c}\ l\ C_l,\quad
\al=\sum_{l=0}^{n_c} \ C_l, \nn\\ \lb{eq:constraints}\\
P(\la)&=&\sum_{l=1}^{n_c}\int d\vlal\ c(l,\vlal)
       \lh\sum_{k=1}^l\de(\laik-\la)\rh\ .\nn
\eea

\noi where $C_l\ev\int d\vlal c(l,\vlal)$. Note that the first constraint
of (\ref{eq:constraints}) need not be imposed separately, since it follows
upon integrating the third constraint over $\la$. Upon introducing
corresponding Lagrange multipliers $\{\hcB$, $\hcP(\la)\}$ to enforce the
constraints, we obtain our order parameter equations. Variation with
respect to the $c(l,\vlal)$ yields

\beq
\ c(l,\vlal)={1\ov l!}\exp\lh-[1+\be(E(l,\vlal)+\hcB+\sum_{k=1}^l
\hcP(\laik))]\rh .
\eeq

\noi
To simplify notation we define $\phi(\la)\ev\exp(-\be\hcP(\la))$ and
$\hB\ev\exp(-(1\e\be\hcB))$:

\beq
c(l,\vlal)={\hB\ov l!}\lh\prod_{k=1}^l\phi(\laik)\rh
\exp(-\be E(l,\vlal))\ .
\eeq

\noi Next we have determine $\hB$ and $\phi(\la)$ by enforcing
the constraints (\ref{eq:constraints}), which leads us to the following
non-linear integral equation for $\phi(\la)$

\bea
\al\phi(\la)&&\!\!\!\lh\sum_{l=1}^{n_c}\int\prod^l_{k=1}(d\laik\phi(\laik))
\de(\la-\la_1){\exp(-\be E(l,\vlal))\ov(l\m1)!}\rh \nn\\ \lb{ie}\\
\times&&\lh\sum_{l=0}^{n_c}\int\prod_{k=1}^l(d\laik\phi(\laik))
{\exp(\m\be E(l,\vlal))\ov l!}\rh^{\m1}\!\!\!=P(\la)\ ,\nn
\eea

\noi We now work out our results for the RHM and RCM  models separately:
\begin{itemize}
\item In the case of the RHM, $E(l,\vlal)=\sum_{k=1}^l
E_1(l,\laik)$, such that (\ref{ie}) reduces to

\bea
\al\phi(\la)&&\lh\sum_{l=1}^{n_c}\exp(-\be E_1(l,\la))
{\cG(\be,l)^{(l\m1)}\ov (l\m1)!}\rh\times\nn\\ \\
&&\times\lh\sum_{l=0}^{n_c}{\cG(\be,l)^l\ov l!}\rh^{\m1}\!\!\!=P(\la)\ ,
\nn
\eea

\noi where $\cG(\be,l)\ev\int d\la_1\phi(\la_1)\exp(-\be E_1(l,\la_1))$\ .
This reduces to a discrete set of equations upon inserting the trial
solution $\phi(\la)=P(\la)[\sum_{l=1}^{n_c}d_l\exp(-\be E_1(l,\la))]^{\m1}
$. The $d_l$ are to be solved numerically from

\beq
d_l=\al{\cG(\be,l)^{(l\m1)}\ov (l\m1)!}\lh\sum_{l'=0}^{n_c}
{\cG(\be,l')^{l'}\ov l'!}\rh^{-1}
\eeq

\noi The results of these calculations are compared with numerical
simulations in FIG. \ref{fg5}, for Gaussian distributed
hydrophilicities/hydrophobicities.

\begin{figure}[t]
\vspace*{5mm}
\setlength{\unitlength}{0.6mm}
\begin{picture}(0,85)
\put( -8, 5){\epsfxsize=140\unitlength\epsfbox{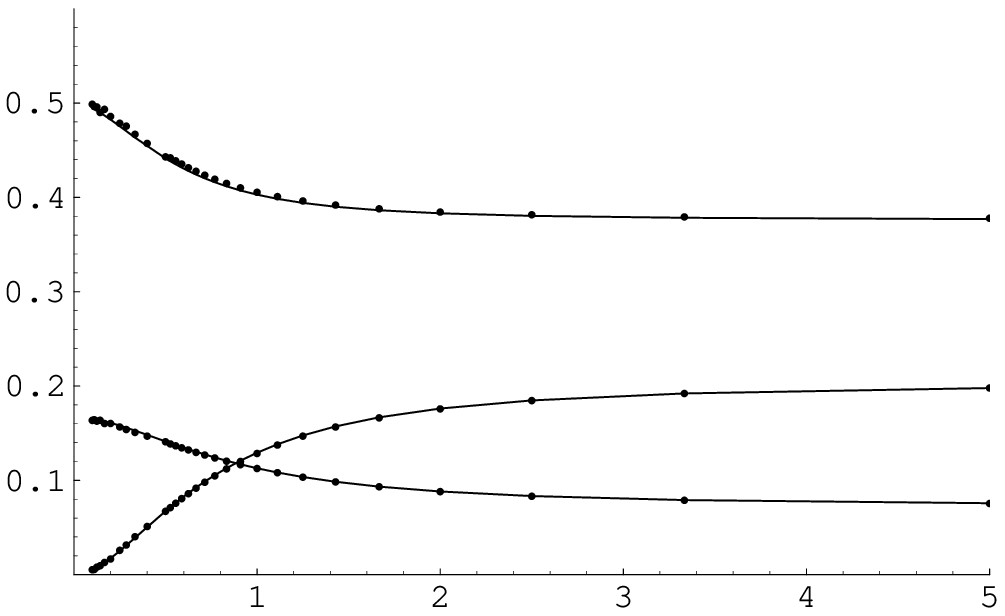}}
\put(112,65){\makebox(1,1){\large $C_1$}}
\put(112,40){\makebox(1,1){\large $C_2$}}
\put(112,25){\makebox(1,1){\large $C_3$}}
\put( 67, 3){\makebox(1,1){\large $T$}}
\end{picture}
\vspace*{3mm}
\caption{
Order parameters $C_l$  as functions of temperature, for the Random
Hydrophobic/Hydrophilic Model (RHM), with $n_c=3$, $\alpha=1$, and
$P(\la)=\cN(0,1)$. Full lines: theory. Dots: numerical simulations
($N=3000$).}
\lb{fg5}
\end{figure}

\item
In the case of the RCM, $E(l,\vlal)={\eps\ov2}(\sum_{k=1}^l\laik)^2$ and
using a Hubbard-Stratonovich transformation, we obtain

\bea
\al\phi(\la)&&\lh\int Dz \exp(\sqrt{\eps\be}\la z)\sum_{l=1}^{n_c}
{\cK(\be,z)^{(l\m1)}\ov(l\m1)!}\rh\times \nn\\
&&\times\lh\int Dz\sum_{l=0}^{n_c}{\cK(\be,z)^l\ov l!}\rh^{\m1}=P(\la)\ ,
\lb{phic}
\eea

\noi in which $Dz\ev dz\exp(-z^2/2)/\sqrt{2\pi}$, and $\cK(\be,z)\!\ev
\!\int d\la_1\phi(\la_1)\exp(\sqrt{-\eps\be}\la_1z)$\ . We are not able to
solve (\ref{phic}) for general $P(\la)$ and/or variable monomer density.
\nl
However, for fixed monomer density, i.e.\nl
$c(l,\vlal)\ev\de_{l,n_c}\lh\prod_{k=1}^l\phi(\laik)\rh
\exp(-\be E(l,\vlal))$, in combination with the choice
$P(\la)=\sqrt{v\ov2\pi}\exp(\m v(\la\m\la_0)/2)$, we obtain

\bea
\phi(\la)&=&\sqrt{a^{(n_c\m1)/n_c}(a\e n_c\eps\be)^{1/n_c}\ov2\pi}\times
\lb{rcc}\\
&\times&\exp(-{a\ov2}\la^2+\la_0(a\e n_c\eps\be)\la-\la_0^2
(a\e n_c\eps\be)/2), \nn\\
a&=&{(v\m n_c\eps\be)+\sqrt{(v\m n_c\eps\be)^2+4v(n_c\m1)\eps\be}\ov2}\ .
\nn
\eea

\noi The results of solving these equations are compared with numerical
simulations in FIG. \ref{fg6}, for Gaussian distributed charges.

\begin{figure}
\vspace*{10mm}
\setlength{\unitlength}{0.56mm}
\begin{picture}(0,150)
\put( 72, 80){\epsfxsize=75\unitlength\epsfbox{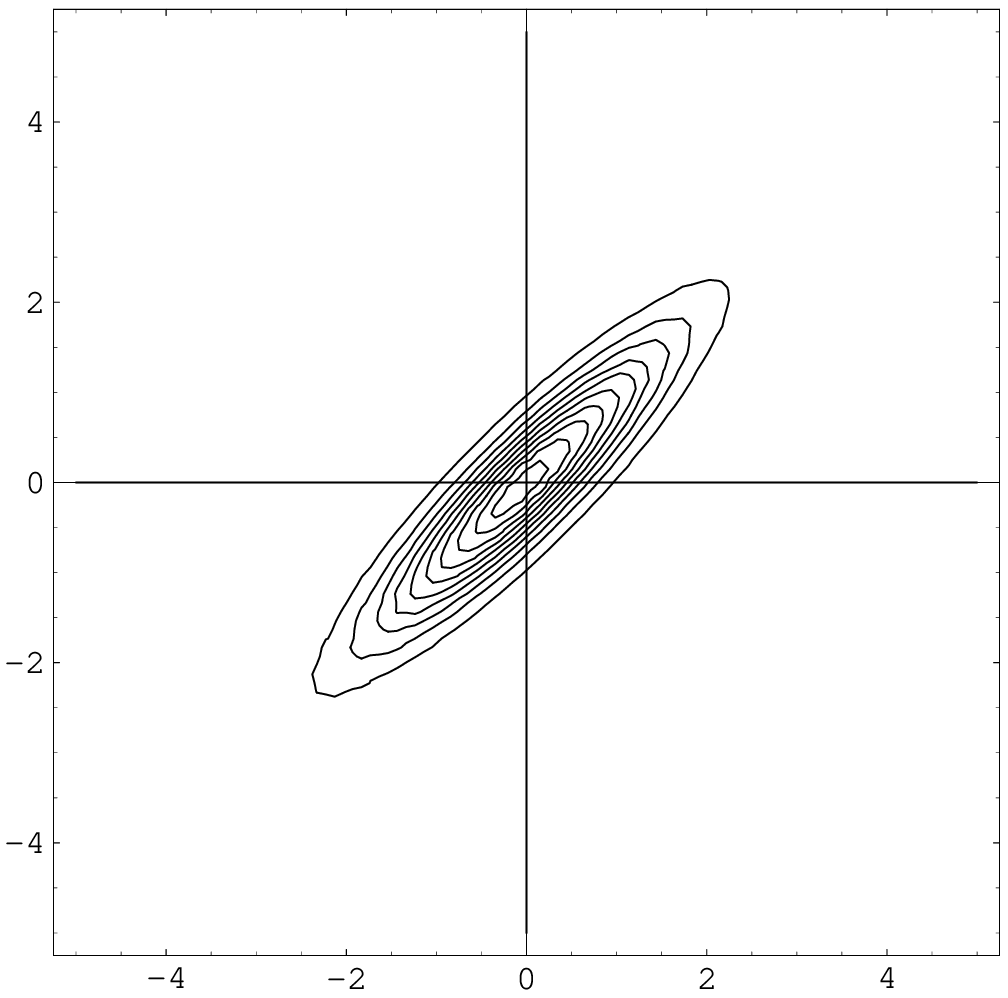}}
\put( -8, 80){\epsfxsize=75\unitlength\epsfbox{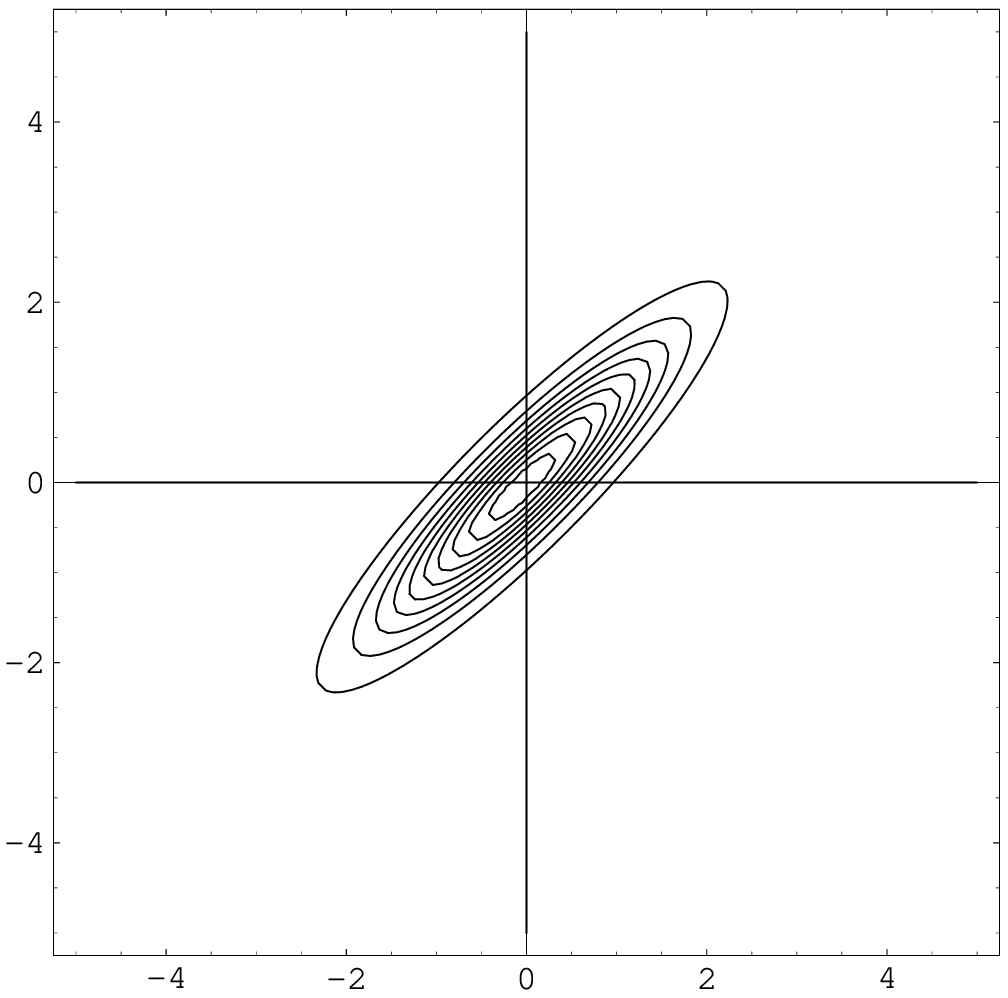}}
\put( 62,122){\makebox(1,1){$\la_1$}}
\put( 34,145){\makebox(1,1){$\la_2$}}
\put(142,122){\makebox(1,1){$\la_1$}}
\put(114,145){\makebox(1,1){$\la_2$}}
\put( 92,150){\makebox(1,1){$\epsilon\!=\!1$}}
\put( 92,145){\makebox(1,1){\rm simulation}}
\put( 12,150){\makebox(1,1){$\epsilon\!=\!1$}}
\put( 12,145){\makebox(1,1){\rm theory}}
\put( 72, 5){\epsfxsize=75\unitlength\epsfbox{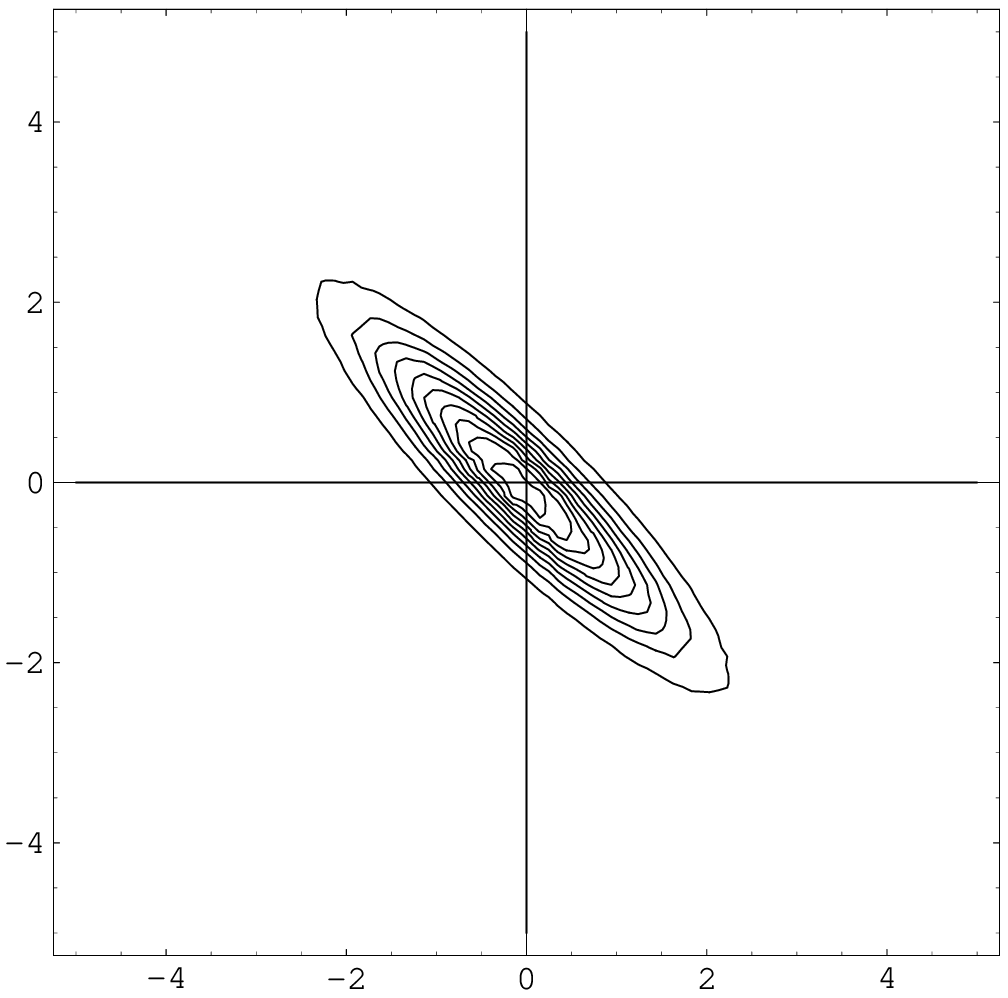}}
\put( -8, 5){\epsfxsize=75\unitlength\epsfbox{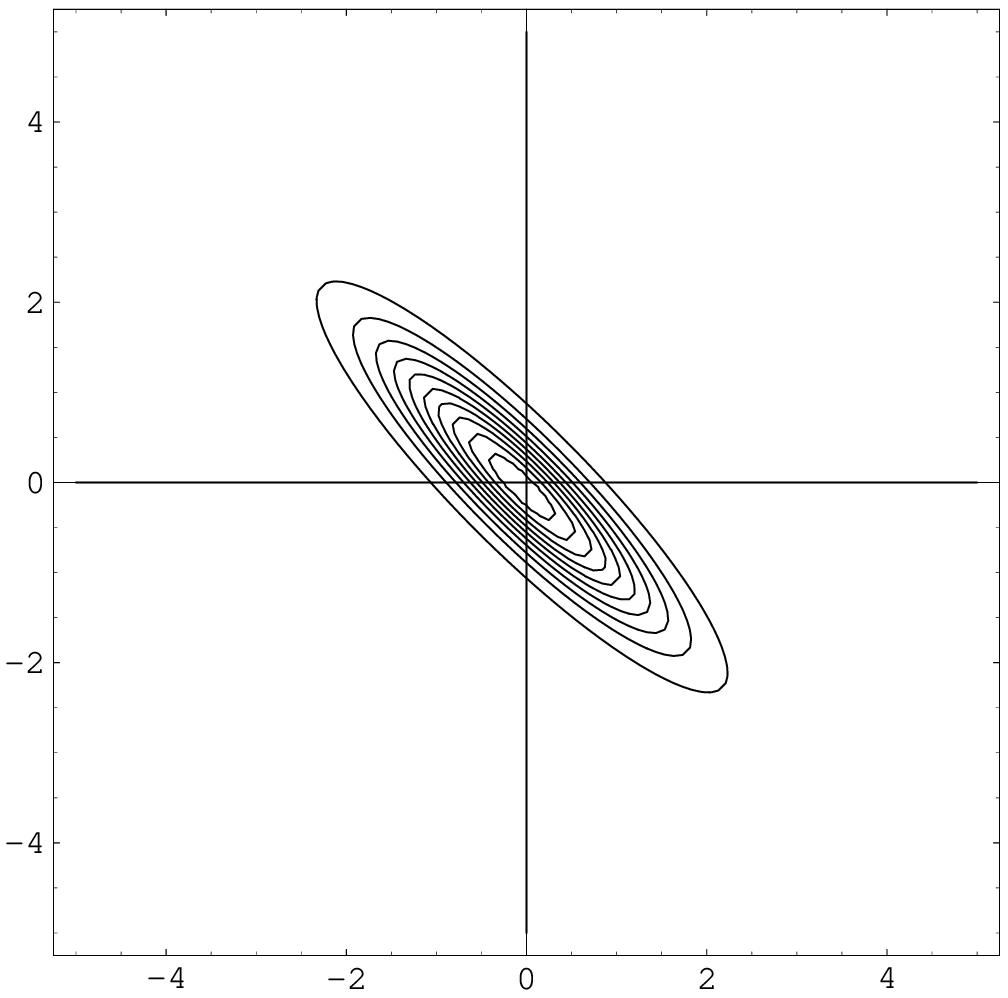}}
\put( 62,47){\makebox(1,1){$\la_1$}}
\put( 34,70){\makebox(1,1){$\la_2$}}
\put(142,47){\makebox(1,1){$\la_1$}}
\put(114,70){\makebox(1,1){$\la_2$}}
\put( 93,75){\makebox(1,1){$\epsilon\!=\!-\!1$}}
\put( 93,70){\makebox(1,1){\rm simulation}}
\put( 13,75){\makebox(1,1){$\epsilon\!=\!-\!1$}}
\put( 13,70){\makebox(1,1){\rm  theory}}
\end{picture}
\caption{
Contour plots of concentrations $c_2(\la_1,\la_2)$ for the RCM with $n_c=2$,
$P(\la)=\cN(0,1)$, $\al=\ha$, $T=0.2$ and $\eps=\pm1$.
Upper graphs: $\epsilon=1$ (equal charges repel).
Lower graphs: $\epsilon=-1$ (equal charges attract).
Left graphs: theoretical  predictions. Right graphs:
numerical simulations ($N=3000$). }
\lb{fg6}
\end{figure}

\end{itemize}

\subsection{RIM}

\noi In the case of the RIM, we can divide the monomers in classes
$\{l,\vlal\}$ where $l$ is the number of monomers and
$\vlal\ev\{\la_{i_k,i_m},k<m=1,..,l\}$. As in the discrete case, we cannot
impose the equivalent of the $P(\lambda)$ condition in
(\ref{eq:constraints}), but at finite temperatures the annealed
approximation is exact (see appendix {\bf B}), and the ensemble
probability that the label vector $\vlal$ occurs, is given by
$\prod_{k<m}^lP(\la_{i_k,i_m})$. Hence, the state-dependent part of free
energy per monomer is

\beq
{\cal F}_{\rm sd}=
\sum_{l=0}^{n_c}\int d\vlal\ c(l,\vlal)\lv E(l,\vlal)+{1\ov\be}
\log({l!\ c(l,\vlal)\ov\prod_{i<k}^lP(\la_{ik})})\rv\ .
\lb{Frei}
\eeq

\noi where $c(l,\vlal)$ is the density of states. The equilibrium
distribution $c^*(l,\vlal)$ is determined by minimization of
(\ref{Frei}), yielding

\beq
c(l,\vlal)=\lh\prod_{i<k=1}^l P(\la_{ik})\rh{\hcN^l\hcB\ov l!}
\exp(-\be E(l,\vlal))
\eeq

\noi where $\hcN$ and $\hcB$ have to be chosen such that the first two
constraints in (\ref{eq:constraints}) are satisfied. The results of solving
the remaining equations numerically are compared with the results of
carrying out numerical simulations in FIG. \ref{fg7}.

\begin{figure}
\vspace*{10mm}
\setlength{\unitlength}{0.6mm}
\begin{picture}(0,85)
\put( -8, 5){\epsfxsize=140\unitlength\epsfbox{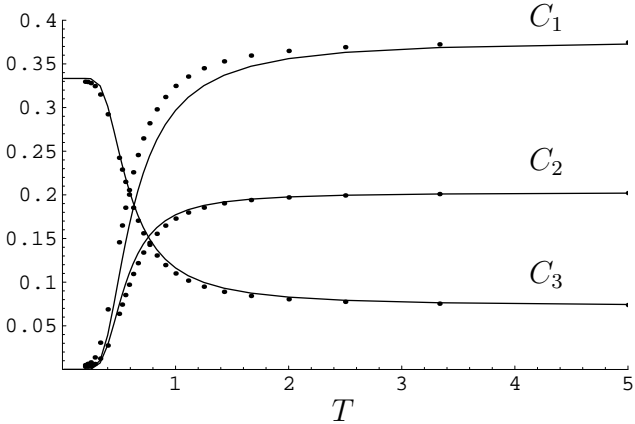}}
\put(112,89){\makebox(1,1){\large $C_1$}}
\put(112,57){\makebox(1,1){\large $C_2$}}
\put(112,32){\makebox(1,1){\large $C_3$}}
\put( 67, 2){\makebox(1,1){\large $T$}}
\end{picture}
\vspace*{5mm}
\caption{
Order parameters $C_\ell$  as functions of temperature, for the Random
Interaction Model (RIM), with $n_c=3$, $\alpha=1$, and $P(\la)=\cN(0,1)$.
Full lines: theory. Dots: numerical simulations ($N=4500$). Deviations are
due to finite size effects, which are more profound in the RIM (which has
bond disorder) than in the RHM and RCM (which have site disorder).}
\lb{fg7}
\end{figure}

\section{Continuous Disorder: Na\"{\i}ve Mean Field Solution  \lb{Con}}

We now compare the solutions of our models, as obtained directly in the
previous section, to the result of following the standard procedure: which
is to calculate the typical quenched free energy via the replica method:

\beq
\lgl-\be F \rgl=\lim_{n\to0}{\lgl\cZ^n\rgl-1\ov n}\ ,
\quad\cZ^n=\Tr_{\{r^a_i\}}\exp\lh-\be\cH\rh\ ,
\lb{Zn}
\eeq

\noi where $\cZ^n$ is the $n$-replicated partition sum and
$\Tr_{\{r^a_i\}}$ denotes the trace over all configurations of the replicas
with the constraint that each box contains at most $n_c$ monomers. We now
perform the average over the disorder, and, in analogy with the continuum
models considered in the literature, we enforce the conditions
$n^a_r\ev\sum_i\de_{r,r^a_i}$ and $q^{ab}_{rr'}\ev\sum_i\de_{r,r^a_i}
\de_{r',r^b_i}$ by introducing the conjugate variables $\n^a_r$ and
$\q^{ab}_{rr'}$. This yields the following expression for the replicated
partition sum:

\bea
\lgl\cZ^n\rgl&=&\sum'_{\{n_r^a\}}\!\int\!\cD\{\n^a_r\}
\sum'_{\{q^{ab}_{rr'}\}}\!\int\!\cD\{\q^{ab}_{rr'}\}\exp\lh
i\sum_{r,a}\n^a_rn^a_r\e\rp\nn \\
&&\hsp{-5mm}\lp+i\sum_{r,r'}\sum_{a<b}\q^{ab}_{rr'}
q^{ab}_{rr'}+g_1(n^a_r,q^{ab}_{rr'})+g_2(\n^a_r,\q^{ab}_{rr'})\rh
\lb{mZn}
\eea

\noi where $g_2$ is given by

\beq
\sum_l\log\lh\Tr_{r^a_l=1}^N\exp(-i\sum_a\n^a_{r^a_l}-
i\sum_{a<b}\q^{ab}_{r^a_lr^b_l})\rh.
\lb{g2}
\eeq

\noi and $g_1$ is given by

\beq
\lc \bay{l}
-\be\la_0(\sum_{ar}n_r^{a2}-nNn_c)+{\be^2\la^2\ov2}
\sum_{ar,br'}n^a_rM_{ar,br'}n^b_{r'}\\ \\
\hsp{63mm}{\rm for\ the\ RHM} \\ \\
-{\eps\be\la_0^2\ov2}\sum_{ar,br'}n^a_rR^{-1}_{ar,br'}n^b_{r'}
-\ha\log(\det(R_{ar,br'}))\\ \\
\hsp{63mm}{\rm for\ the\ RCM} \\ \\
{-\be\la_0\ov2}(\sum_{ar}n_r^{a2}-nN)+{\be^2\la^2\ov4}\sum_{ar,br'}
(M_{ar,br'})^2 \\ \\
\hsp{63mm}{\rm for\ the\ RIM}
\eay\rp
\lb{g1}
\eeq

\noi where

\bea
M_{ar,br'}&=&n^a_r\de_{r,r'}\de_{a,b}+q^{ab}_{rr'}(1-\de_{a,b})\ ,\nn\\
\lb{mat}\\
R_{ar,br'}&=&(1\e\eps\be\la^2n^a_r)\de_{r,r'}\de_{a,b}+
           {\eps\be\la^2\ov2}q^{ab}_{rr'}(1-\de_{a,b}) \nn
\eea

\noi The MF treatment up to equation (\ref{mat}) is perfectly valid.
At this stage, however, it is important to note that the
$n^a_r\in\{0,1,..,n_c\}$ and $q^{ab}_{rr'}\in\{0,1,..,n^a_rn^b_{r'}\}$ are
discrete microscopic variables, rather than self-averaging order parameters
(compare to e.g. spin-glass theory \cite{MPV}). Of these at finite density,
there are $\cO(N)$ and $\cO(N^2)$ respectively, while in the finite
dimensional analogon, the volume occupied by the polymer is $\cO(N)$. Note,
furthermore, that for any $n_c\simeq\cO(N^\nu)$ with $\nu<1$ (in our case
$\nu=0$), even in the most compact configuration the vast majority of the
$q^{ab}_{rr'}$ will have to be zero. In appendix {\bf B}, we will show that
at finite $T$ the non-zero $q^{ab}_{rr'}$'s will be 1. Having a finite
fraction of the non-zero $q^{ab}_{rr'}$'s $>1$, leads to a decrease in the
entropy of order $N\log(N)$, while the energetic gain is at most extensive
(this is specific to our $\infty$-dimensional model). Hence, the integrals
over the $\n^a_r$ and $\q^{ab}_{rr'}$ cannot be treated as SP integrals
(there is no large parameter $N$).
Nevertheless, since the integrals cannot be performed explicitly, one
could consider SP integration as a first approximation.
We here find that a SP treatment of the $\n^a_r$ leads to at least
qualitatively correct results, which may explain the relative success of
MF models for homo-polymers \cite{DE}. In many studies the $n^a_r$ are
kept constant anyway, as one focuses on the properties of the compact
phase, eliminating the SP-problem altogether. We will show, however, that
a (na\"{\i}ve) saddle-point treatment of the $\q^{ab}_{rr'}$ may lead to
qualitatively incorrect results.
Several variational approaches have been proposed to deal with this problem
\cite{GLO}-\cite{SG}. The quality of a variational approach depends on the
quality of the trial function(s), and there is always the risk that the
true physical behaviour is not included.
To develop a self-consistent MF theory for random hetero-polymers, it may
be necessary to find a good set of ``true'' order parameters. It may well
be that the origin of some of the incongruences between the results
obtained by analytical methods for continuum models and those found
numerically for lattice models \cite{TvMM,GK}, lay in the application of
the saddle-point method for strongly fluctuating quantities such as the
$q^{ab}_{rr'}$ and $\q^{ab}_{rr'}$.

\subsection{Mean Field solution of the RHM \lb{MFH}}

\noi In this model, one can completely avoid the $q^{ab}_{rr'}$ (and hence
the $\q^{ab}_{rr'}$), by using the equality

\beq
\sum_{rr'}n^a_rq^{ab}_{rr'}n^b_{r'}=\sum_in^a_{r^a_i}n^b_{r^b_i}\ .
\lb{sh}
\eeq

\noi Due to the fact that the $q^{ab}_{rr'}$ appear linearly in $g_1$
(\ref{g1}), the introduction of the $\q^{ab}_{rr'}$ and the solution of the
SP equations, replaces the $q^{ab}_{rr'}$ by their average value.
As they are summed over, one obtains the same (exact) result as by using
(\ref{sh}) directly. Since only one-replica-index parameters remain, no
replica symmetry breaking (RSB) will occur. After some rewriting, using the
Hubbard-Stratonovich transformation, replica symmetry (RS), and keeping
only leading terms in $n$, we obtain for $g2$

\beq
nN\int Dz\log\lh\Tr_{r'=1}^N
\exp(\m i\ \n_{r'}\m{\be^2\la^2\ov2}n^2_{r'}\e\be\la z n_{r'})\rh\ .
\lb{g2h}
\eeq

\noi Similarly, the SP equations with respect to the $\n^a_r$,
can be written

\beq
n_r=N\int Dz{\exp(-i\n_r   -{\be^2\la^2\ov2}n^2_r   +\be\la z n_r   )\ov
\Tr_{r'=1}^N \exp(-i\n_{r'}-{\be^2\la^2\ov2}n^2_{r'}+\be\la z n_{r'})}\ ,
\lb{nrh}
\eeq

\noi which have to interpreted as equations for the $\n^a_r$. As explained
before, for any finite $n_c$, the most general solution for the $n_r$ can
be written down in terms of a finite number of parameters $C_l$. With the
definitions

\bea
&&\De_k\ev{\sum_r (n_r)^k\ov N}=\sum_{l=1}^{n_c}C_ll^k=
\sum_{l=1}^{n_c}f_ll^{(k-1)}\ ,\quad\fa k\in I\!\!N\ ,\nn\\
&&h(z,l)\ev-i\n_l\m\ha\be^2\la^2l^2\e\be\la zl,\lb{De} \\
&&\cN(z)\ev\sum_{l=0}^{n_c}C_l\exp(h(z,l))\ ,\nn
\eea

\noi and adding the entropic correction (\ref{ent}), we obtain the
following expression for the state-dependent part of the free energy per
particle in the SP approximation

\bea
\cF_{\rm sd}&=&\m\la_0(n_c\m\De_2)\m{\be\la^2\ov2}\De_3-
{1\ov\be}\lh i\sum_{l=1}^{n_c}lC_l\n_l+\rp\nn\\
&&~~~\lp+\int Dz \log(\cN(z))\m\sum_{l=0}^{n_c}C_l\log(C_l)\rh\ ,
\lb{Fqh}
\eea

\noi With the constraint $\De_1=1$, the equations (\ref{nrh}) reduce to
$n_c\e1$ non-equivalent equations for the $\n_l$, and $n_c\m1$ equations
for the $C_l\ (l=2,..,n_c)$:

\bea
&&l=\int Dz\exp(h(z,l))/\cN(z)\ ,
\lb{lh}\\
&&\la_0(l\m1)\m{\be\la^2\ov2}(l^2\m1)\e{1\ov\be}\lh i(\n_1\m\n_l)\e{1\ov l}
\log(C_l)-\rp\nn\\
&&~~~~~~~~~~~~~~~~\lp-\log(C_1)\e{(l\m1)\ov l}\log(C_0)\rh=0\ .
\lb{flh}
\eea

\noi From (\ref{lh}), one sees that for all temperatures $-i\ \n_0=-\infty$,
and that only differences between the $\n_l$ are important. Thus, the
$\n_l$ are defined up to a constant, which drops out of the final
expression for the free energy. At any temperature, the free energy
(\ref{Fqh}) depends of a finite number of parameters $C_{n_c},..,C_2$ and
$\n_{n_c},..,\n_2$ ($C_1$ is fixed by normalisation and $\n_1$ fixes an
irrelevant gauge). The SP values of these parameters have to be
obtained numerically from (\ref{lh}) and (\ref{flh}). The results we
obtain, are qualitatively correct, but as explained in appendix {\bf A} one
can do better by adding the high temperature correction

\beq
{1\ov\be}\sum^{n_c}_{l=2}C_l\log({l!\ov l^l})
\lb{cor}
\eeq

\noi to the free energy (\ref{Fqh}) (and make the corresponding correction
in (\ref{flh})), to obtain even quantitatively excellent results.
\nl
To show that we obtain the correct low temperature behaviour, we solve
equations (\ref{lh},\ref{flh}) (with the correction) in the limit
$\be\to\infty$ analytically. It can be shown that in that limit the $C_l$
for intermediate values $1\!<\!l\!<\!n_l$, vanish. Hence, we immediately
start with this simple ansatz: $C_{n_c}N$ boxes contain $n_c$ monomers,
$C_1N=(1\m n_cC_{n_c})N$ boxes contain 1 monomer, and
$C_0N\ev(\al\m1\e(n_c\m1)C_{n_c})N$ boxes contain no monomers. We call this
{\em phase separation}, because the monomers are either in the maximally
compact or in the swollen phase.
\nl
In the limit $\be\to\infty$ only the leading exponent survives, such that

\bea
i\ \n_c&=&-\ha\be^2\la^2n_c^2 +\be\la An_c   +C\ ,\nn\\
i\ \n_1&=&-\ha\be^2\la^2\ \ \ +\be\la A\ \ \ +C\ .
\lb{nps}
\eea

\noi where $A$ is defined via

\beq
f_{n_c}=n_cC_{n_c}=\int^\infty_A Dz=\ha \erfc({A\ov\sqrt{2}})\ ,
\lb{fc}
\eeq

\noi Inserting all this into (\ref{Fqh}), we obtain for the ground state
energy

\bea
E_0(A^*)&=&-(n_c\m1)\lh\la_0\ (1\m f^*_{n_c})+\la{\exp(\m A^{*2}/2)\ov
\sqrt{2\pi}}\rh\nn\\
&=&-(n_c\m1)\int^\infty_0 d\la\ \la P(\la)\ ,
\lb{E0h}
\eea

\noi where $A^*$ is the value that maximizes (\ref{E0h}), i.e.

\beq
A^*={\la_0\ov\la},\hsp{1cm} f_{n_c}^*=\ha\erfc({\la_0\ov\sqrt{2}\la})\ .
\lb{Ah}
\eeq

\noi This result makes perfect sense:
given $\la_0$ and $\la$, the fraction of monomers that are negative, i.e.
hydrophobic, is exactly $f_{n_c}^*$. For energetic reasons, these monomers
want to be in the maximally compact state ($n_{r_i}=n_c$), while the others
are hydrophilic and want to be maximally swollen ($n_{r_i}=1$). Note that
the free energy can become arbitrarily negative for $n_c\to\infty$.
At high temperature we know the exact solution, and at low temperature
the energetic term dominates and is calculated exactly by SP
integration, because $\be$ acts as the large parameter. Numerical analysis
shows that also at intermediate temperatures the obtained curves are
indistinguishable from the exact solution (FIG.\ \ref{fg5}).
\nl
We conclude that the RHM is probably the only model in which the SP method
leads to qualitatively correct results. This could be anticipated by the
observation that using (\ref{sh}), the $q^{ab}_{rr'}$ could have been
avoided altogether.

\subsection{Mean Field Solution of the RCM \lb{MFC}}

\noi Before turning to the technical analysis of the model, we anticipate
what to expect for the low temperature behaviour.
\begin{itemize}
\item
First, we consider the case $\eps=+1$, such that equal charges repel each
other. When the average charge is $0$ ($\la_0=0$), we expect the monomers
to group together into boxes such that the total charge in each box goes
to zero, because the energy is exactly the sum of the squares of the local
charges $\la_r$ (\ref{HC}). When $\la_0\neq0$ (take $\la_0>0$ without loss
of generality), however, there is a total charge $N\la_0$, a total negative
charge $N\la_n$, and we define $A$ by

\beq
\int_A^\infty d\la_iP(\la_i)(\la_i-A)=-\la_n\ .
\lb{Ac}
\eeq

\noi The ground state energy per monomer $E_0$ is then limited from below
by

\beq
E_0\geq\int_0^Ad\la_iP(\la_i)\la^2_i+A^2\int^\infty_Ad\la_iP(\la_i)\ .
\lb{E0cp}
\eeq

\noi In this case the most positive charges are compensated by negative
ones to become $A$, until all negative charges are used up (with a finite
number of charges in each box, it is not guaranteed that this can indeed be
done). The remaining (excess) positive charges will stay alone in a box.
A closer look at the expression for $g_1$ (\ref{g1}) shows that it is well
defined at all temperatures for $\eps=+1$ (equal charges repel).
\item
For the case $\eps=-1$, we expect the monomers to group together in order
to maximize $|\la_r|$ for all the occupied boxes, and the ground state energy
per monomer is hence limited by

\beq
E_0\geq-\int_{-\infty}^\infty d\la_i P(\la_i)(n_c\la_i)^2\ .
\lb{E0cm}
\eeq

\noi
A closer look at the expression for $g_1$ (\ref{g1}) reveals that there is
a critical $\be_c$ (dependent on $n_c,\la^2$, but independent of $n$) for any
finite $n$, such that for $\be\geq\be_c$ the matrix $R_{ar,br'}$ is no
longer positive definite. This is of course an artifact of the replica
method, which first calculates the annealed average of $n$ copies of the
system and then takes the limit $n\to0$. For any finite $n>0$, there is a
phase transition at $\be_c$ after which the variance of the $\la_i$
diverges. The central problem is the fact that this phase transition makes
the result for finite $n$ useless to obtain information about the system in
the limit $n\to0$. Therefore, the replica method is not applicable for low
temperature results with $\eps=\m1$. This problem should not occur for
distributions $P(\la)$ with the property that $\lim_{\la\to\pm\infty}P(\la)
\exp(a\la^2)\to0\ ,\quad \fa a\in I\!\!R$.
\end{itemize}
We now follow the same procedure as in the previous section.
The SP equations for the $q^{ab}_{rr'}$ yield

\beq
-i\ \q^{ab}_{rr'}={\be^2\la^2\la_0^2\ov2}N^a_rN^b_{r'}\ ,\hsp{1cm}
N^a_r\ev{n^a_r\ov(1+\eps\be\la^2n^a_r)}\ ,
\lb{qc}
\eeq

\noi such that (at least in the SP approximation), the
two-replica parameters reduce to a product of one-replica order
parameters, and no RSB will occur. As explained before, for any finite
$n_c$ the most general solution for the $n_r$ can be written down in
terms of a finite number of parameters $C_l$. With the definitions
(\ref{De}) and

\bea
h(z,l)\ev-i\ \n_l-{\be^2\la^2\la_0^2\ov4}N^2_l+\be\la\sqrt{\la_0^2\ov2}z
N_l\ ,\lb{hc} \\
\cN(z)\ev\sum_{l=1}^{n_c}C_l\exp(h(z,l))\ ,\quad
N_l\ev{l\ov(1+\eps\be\la^2l)},
\lb{Nc}
\eea

\noi after some rewriting, using result (\ref{qc}), RS, the entropic
correction (\ref{cor}) and the Hubbard-Stratonovich transformation, we
obtain for the state-dependent part of free energy

\bea
\cF_{\rm sd}&=&{\eps\la^2_0\ov2}\sum_l lC_lN_l\m{1\ov\be}\lv
       i\sum_{l=1}^{n_c}lC_l\n_l\e\int Dz \log(\cN(z))-\rp\nn\\
&&\lp-\sum_{l=0}^{n_c}C_l\log({l!C_l\ov l^l})\e
\ha\sum_{l=1}^{n_c}C_l\log\lh{N_l\ov l}\rh\rv\ .
\lb{fcq}
\eea

\noi The $\n_l$ have to be determined from the fixed point equations

\beq
l=\int Dz{\exp(h(z,l))\ov\cN(z)}\ ,\lb{nrc}
\eeq

\noi and we obtain the result that the $q_{jk}\simeq\cO(1/N)$. Again, the
$q_{jk}$ are replaced by their average values, but in contrast to the RHM,
this has severe consequences. In the low temperature limit we find that
$N_l\simeq1/\eps\be\la^2$, and $-i\n_l=\log(l)$ $\fa l$, such that we
obtain

\beq
E_0=0
\eeq

\noi for the ground state energy. Because of (\ref{E0cp}), this result can
obviously not be correct for $\la_0\neq0$.

\subsection{Mean Field Solution of the RIM \lb{MFI}}

\noi Following a similar procedure for the RIM, it is not possible to
decouple the ``saddle-point'' equations with respect to the
$\q^{ab}_{rr'}$ for the different replica's, because the $q^{ab}_{rr'}$
are not the product of single replica quantities as was the case for the
RHM and RCM. Therefore, extra approximations, such as variational
approaches, are needed. Nevertheless, one can see that in a SP treatment,
the $q^{ab}_{rr'}$ will again be replaced by their average values. From
(\ref{g1}) one easily sees that the $q^{ab}_{rr'}$ dependent term is
$\cO(N)$ when $q^{ab}_{rr'}\in\{0,1,..,n_c\}$, but becomes $\cO(1)$ when we
replace them by an averaged value.
\nl
In appendix {\bf B}, we show that at any finite temperature the dominant
contribution comes from the entries $q^{ab}_{rr'}=1$\ . This alternatively
explains why the $q$ dependent part of (\ref{g1}) for the RIM will be a
constant and the quenched average reduces to the annealed one at finite
$T$. Furthermore, we introduce a set of true order parameters which can
be treated using the SP method. We show that a finite fraction of
$q^{ab}_{rr'}>1$ implies a reduction of the entropy of order
$\cO(N\log(N))$, while the energetic gain is at most extensive. Only at
reduced temperatures $T'=T/\sqrt{\log(N)}$, the system starts deviating
from the annealed behaviour, because the energetic part of the free energy
can compete with the over-extensive part of the entropy. We only signal the
onset of this phenomenon. A detailed study of the low temperature behaviour
is beyond the scope of this present paper, and has been done for the very
similar matching problems \cite{VM,MP,N}. Hence, due to the invariance of
the model for permutations of the monomers (which is definitely not true
for finite dimensional models!), it turns out that the simple annealed
average is equivalent to the quenched one (at finite temperatures). We will
return to this issue in the next section on annealed averages.
\nl
Nevertheless, even at finite temperatures deviations from the annealed
behaviour are observed in the simulations (see FIG. 6). These are finite
size effects, because limitations in computer memory do not allow system
sizes large enough to make $1/\sqrt{\log(N)}$ effects negligible.
\nl

\section{Continuous Disorder: Grand Ensemble Approach}

\noi An alternative approximation scheme is the Grand Ensemble Approach.
It consists in approximating the quenched average by a series of
constrained annealed averages, and is originally due to Morita \cite{Mo}.
For a review and detailed description, we refer to \cite{Ku}.
Since the chain constraint is negligible in our simple model, we expect
good results upon fixing only the moments of the overall probability
distribution of the randomness. In finite dimensional models, however,
this does not prevent the random variables to choose their most favorable
position along the chain. Therefore, one may have to enforce conditions on
the correlations along the chain, in order to obtain a good approximation
for the quenched free energy \cite{TvMM}. Apart from being an approximation
for the quenched case, the (constrained) annealed average also describes a
different physical situation where the monomers are allowed to change their
properties on the same time scale as their configuration.
\nl
Using the same notation as in the previous section, we find that the
equations for the $\n_l$ are temperature independent, and the same as
in the quenched case at infinite temperature. Therefore, we can correct
them exactly to yield the entropic term

\beq
\chi\ev\sum_{l=0}^{n_c}C_l\log(l!\ C_l)\ .
\lb{chi}
\eeq

\subsection{Annealed Averages for the RHM}

\noi First, we calculate the average of the partition sum directly

\beq
\exp\lh-\be N \cF_{a_0}\rh\ev\lgl\cZ\rgl\ ,
\lb{mZ0}
\eeq

\noi which yields the annealed free energy

\beq
\cF_{a_0}=\m\la_0(n_c\m\De_2)\m\ha\be\la^2(\De_3\m2n_c\De_2\e n_c^2)+
{\chi\ov\be}\ .
\lb{Fa0}
\eeq

\noi At intermediate temperatures, the SP equations for the
$f_l,l=2,..,n_c$ have to be solved numerically. In the low temperature
limit however, they can be solved analytically with the phase separation
ansatz ($f_1=1-f_{n_c}$), and yield for the ground state energy

\beq
E^{a_0}_0=(1\m f^*_c)(n_c\m1)\lv\la_0\e\ha\be\la^2(n_c\m1)\rv.
\lb{Ea0}
\eeq

\noi The minimum occurs clearly for $f^*_c=0$, such that all the monomers
will typically be hydrophilic and alone in their boxes.
\nl
In FIG.\ \ref{fg8}.\ {\bf a}, we observe re-entrant cross-over from a
swollen via a compact back to a swollen phase for sufficiently negative
$\la_0$ , indicating real re-entrant phase transitions in models with an
extensive entropy \cite{TvMM}.
\nl
\nl
Secondly, we calculate the average of the partition sum imposing the
constraint that the average of the $\la_i$ is equal to that of the
quenched one

\beq
\exp\lh-\be N \cF_{a_1}\rh\ev\lgl\cZ\ \ \de(\sum_i\la_i\m\la_0N)\rgl.
\lb{mZ1}
\eeq

\noi We introduce the conjugate variable $\hM$ to enforce the constraint
on the average. This is a genuine self-averaging order parameter with
with SP value $\hM=\be(\De_2-n_c)$. The free energy is then
given by

\beq
\cF_{a_1}=-\la_0(n_c-\De_2)-\ha\be^2\la^2(\De_3-\De^2_2)+{\chi\ov\be}\ .
\lb{Fa1}
\eeq

\noi At intermediate temperatures, the SP equations for the
$f_l,l=2,..,n_c$ have to be solved numerically. In the low temperature
limit, however, they can be solved analytically with the phase separation
ansatz ($f_1=1-f_{n_c}$) to yield for the ground state energy

\beq
E^{a_1}_0=\m\la_0(n_c\!\m1)\e\ha\be\la^2(n_c\!\m1)^2(f_{n_c}\m f_{n_c}^2)\ .
\lb{Ea1}
\eeq

\noi The minimum is at $f^*_c=\ha$, such that half of the monomers will be
hydrophobic and in a completely filled

\begin{figure}
\setlength{\unitlength}{0.6mm}
\begin{picture}(0,265)
\put( -6,185){\epsfxsize=140\unitlength\epsfbox{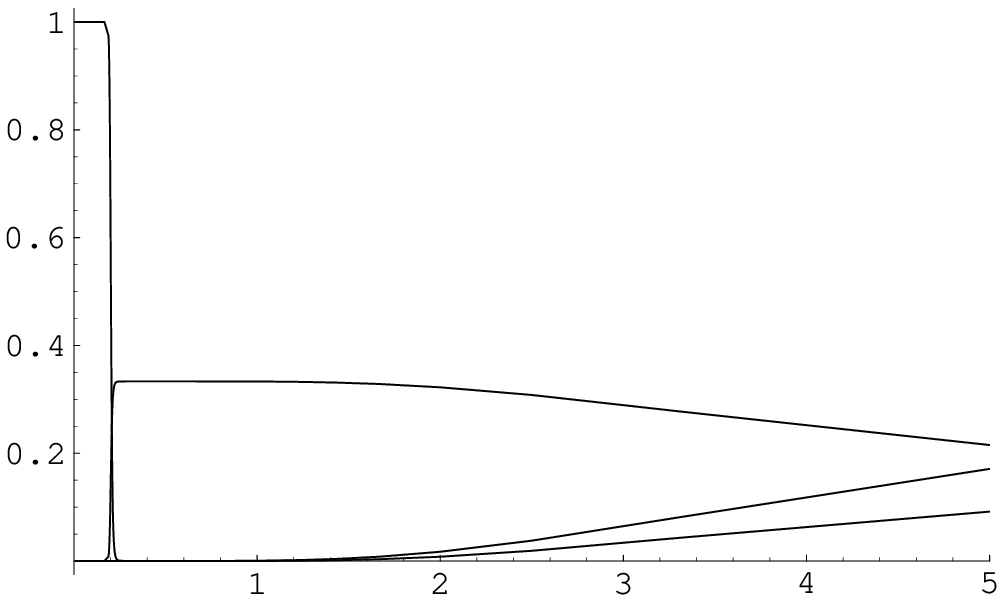}}
\put( 14,260){\makebox(1,1){${\rm\bf a)}$}}
\put( 69,185){\makebox(1,1){$T$}}
\put(138,205){\makebox(1,1){$C_2$}}
\put(132,213){\makebox(1,1){$C_3$}}
\put( 14,240){\makebox(1,1){$C_1$}}
\put( -6, 95){\epsfxsize=140\unitlength\epsfbox{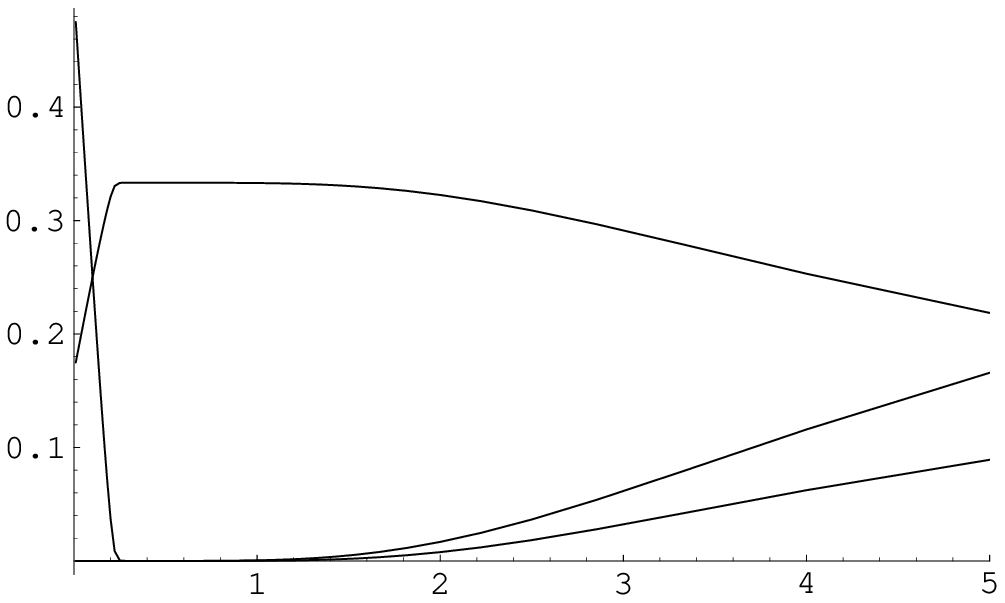}}
\put( 14,170){\makebox(1,1){${\rm\bf b)}$}}
\put( 69, 95){\makebox(1,1){$T$}}
\put(114,117){\makebox(1,1){$C_2$}}
\put(114,127){\makebox(1,1){$C_1$}}
\put(114,146){\makebox(1,1){$C_3$}}
\put( -8,  5){\epsfxsize=140\unitlength\epsfbox{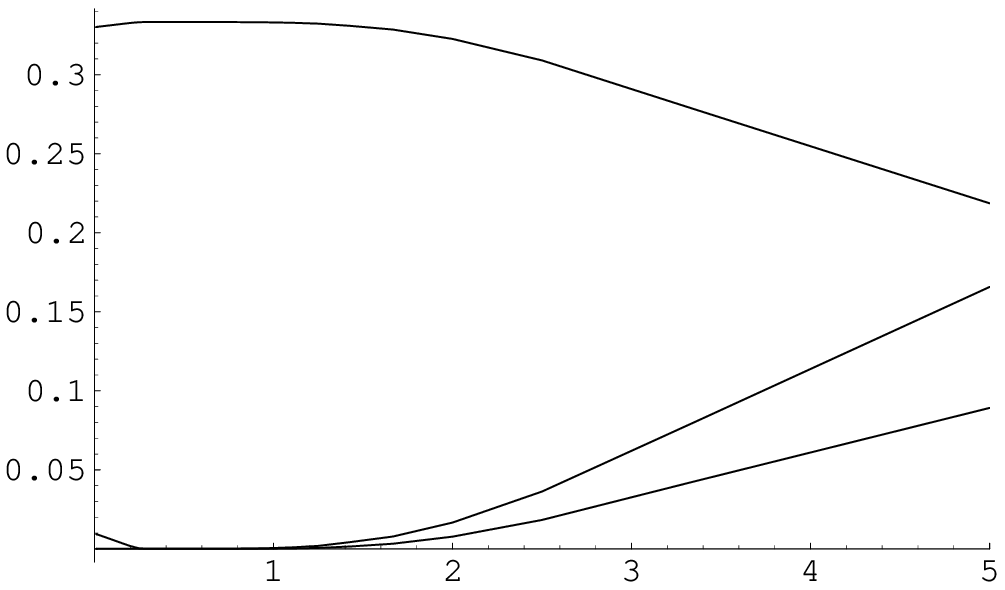}}
\put( 12, 75){\makebox(1,1){${\rm\bf c)}$}}
\put( 67,  5){\makebox(1,1){$T$}}
\put(112, 32){\makebox(1,1){$C_2$}}
\put(112, 47){\makebox(1,1){$C_1$}}
\put(112, 71){\makebox(1,1){$C_3$}}
\end{picture}
\caption{ {\bf a,b,c}\quad
The order parameters $C_l$ as functions of temperature, for the Random
Hydrophilic/Hydrophobic model (RHM), with $n_c=3$, $\al=1$, and
$P(\la)=\cN(-5,1)$: {\bf a)} Simple annealed, {\bf b)} annealed with fixed
mean, {\bf c)} annealed with fixed mean and variance.}
\lb{fg8}
\end{figure}

\noi box, while the other half will be
hydrophilic and alone in their box.
In FIG.\ \ref{fg8}.\ {\bf b}, we observe cross-over to phase separation at
low temperatures, indicating a real phase transition in models with an
extensive entropy \cite{TvMM}.

\noi Thirdly, we impose the constraints that both the average and the
variance of the $\la_i$ must be equal to the quenched ones.

\beq
\exp\lh-\be N\cF_{a_2}\rh\ev\llgl\!\cZ\de(\sum_i(\la_i\m\la_0)
\de(\sum_i((\la_i\m\la_0)^2\m\la^2)\!\rrgl .
\lb{mZ2}
\eeq

\noi To impose these constraints, we introduce the conjugates variables
$\hM$ and $\hS$. Their SP values are
$\hM={\la_0\hS\ov\la^2}\e\be(\De_2\m n_c)$, and
$\hS=\ha(\m1\e\sqrt{1\e4\la^2\be^2(\De_3\m\De_2^2)})$.
The free energy is then given by

\beq
\cF_{a_2}\!\!=\m\la_0(n_c\m\De_2)\e{\log(1\e\hS)\ov2\be}-
{\hS\ov2\be}-{\be\la^2\ov2(1\e\hS)}(\De_3-\De_2^2)+{\chi\ov\be}\ .
\lb{Fa2}
\eeq

\noi At intermediate temperatures, the SP equations for the
$f_l,l=2,..,n_c$ have to be solved numerically. In the low temperature
limit, however, they can be solved analytically with the phase separation
ansatz ($f_1=1-f_{n_c}$) to yield for the ground state energy

\beq
E^{a_2}_0=\la_0(n_c-\De_2)-\la\sqrt{\De_3-\De_2^2}\ .
\lb{Ea2}
\eeq

\noi The minimum occurs for

\beq
f^*_c=\ha(1-{\la_0\ov\sqrt{\la^2+\la_0^2}})\ ,
\lb{fc2}
\eeq

\noi which is already very close to the quenched result (\ref{Ah}). In
FIG.\ \ref{fg8}.\ {\bf c} we observe that the cross over type of behaviour
in the case of very negative $\la_0$ has vanished as it should for the
quenched case. In FIG.\ \ref{fg9}, we compare approximation (\ref{Fa2})
with quenched numerical simulations.

\begin{figure}
\setlength{\unitlength}{0.6mm}
\begin{picture}(0,90)
\put( -8, 5){\epsfxsize=140\unitlength\epsfbox{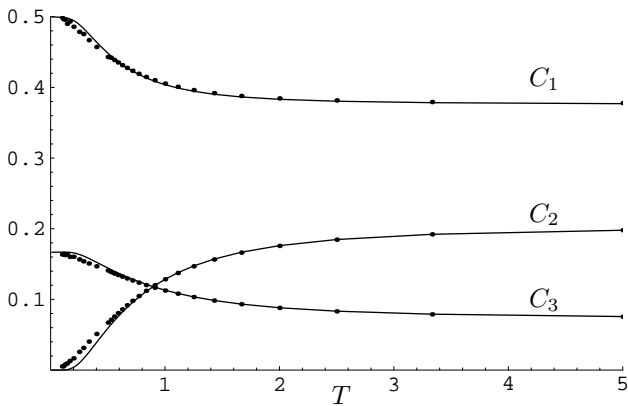}}
\put(112,75){\makebox(1,1){$C_1$}}
\put(112,45){\makebox(1,1){$C_2$}}
\put(112,26){\makebox(1,1){$C_3$}}
\put( 67, 5){\makebox(1,1){$T$}}
\end{picture}
\caption{
The order parameters $C_l$ as a function of temperature, for the Random
Hydrophilic/hydrophobic Model (RHM) with $n_c=3$, $\al=1$, and
$P(\la)=\cN(0,1)$. Full lines: annealed theory with fixed mean and
variance. Dots: numerical (quenched) simulations ($N=3000$). The results
are qualitatively correct.}
\lb{fg9}
\end{figure}

\subsection{Annealed Averages for the RCM}

\noi In order to describe both the cases $\eps=\pm1$ at all temperatures,
we have limited ourselves to the case where both the average and the
variance of the disorder are constrained. Introducing the conjugate
variables $\hM$ and $\hS$ respectively, we obtain the following expression
for the free energy

\bea
\cF_{a_2}\!\!&=&\!-\ha\lv\hS\m\log(1\e\hS)\e{\la^2\hM^2\ov(1\e\hS)}\m
\sum_{l=0}^{n_c}C_l\log(1\e{\eps\be\la^2l\ov(1\e\hS)})\rp\nn\\
&-&\lp\eps\be(\la_0\e{\hM\la^2\ov(1\e\hS)})^2\sum_{l=0}^{n_c}C_l
{l^2\ov(1\e{\eps\be\la^2l\ov(1\e\hS)})}\rv\e{\chi\ov\be}\ .
\lb{Fa2c}
\eea

\noi This approximation is compared with the results of quenched numerical
simulations in FIG. 10.{\bf a,b}. One observes that this leads to at least
qualitatively correct results.

\begin{figure}
\setlength{\unitlength}{0.6mm}
\begin{picture}(0,180)
\put(  7,165){\makebox(1,1){${\rm\bf a)}$}}
\put(112,116){\makebox(1,1){$C_3$}}
\put(112,131){\makebox(1,1){$C_2$}}
\put(112,153){\makebox(1,1){$C_1$}}
\put( 77, 95){\makebox(4,1){$T$}}
\put( -8, 95){\epsfxsize=140\unitlength\epsfbox{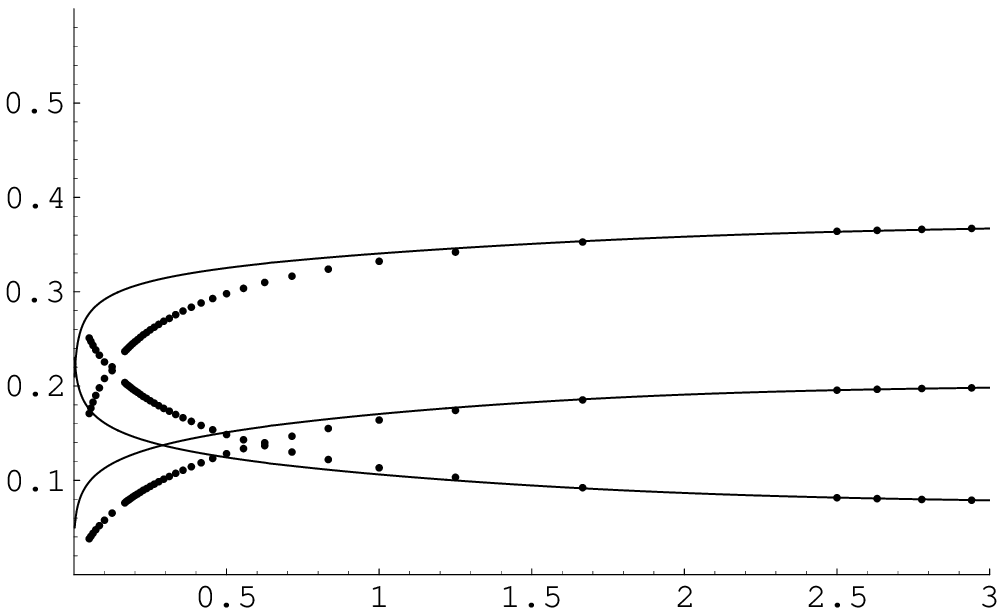}}
\put(  7, 80){\makebox(1,1){${\rm\bf b)}$}}
\put(112, 25){\makebox(1,1){$C_3$}}
\put(112, 41){\makebox(1,1){$C_2$}}
\put(112, 64){\makebox(1,1){$C_1$}}
\put( 77,  5){\makebox(4,1){$T$}}
\put( -8,  5){\epsfxsize=140\unitlength\epsfbox{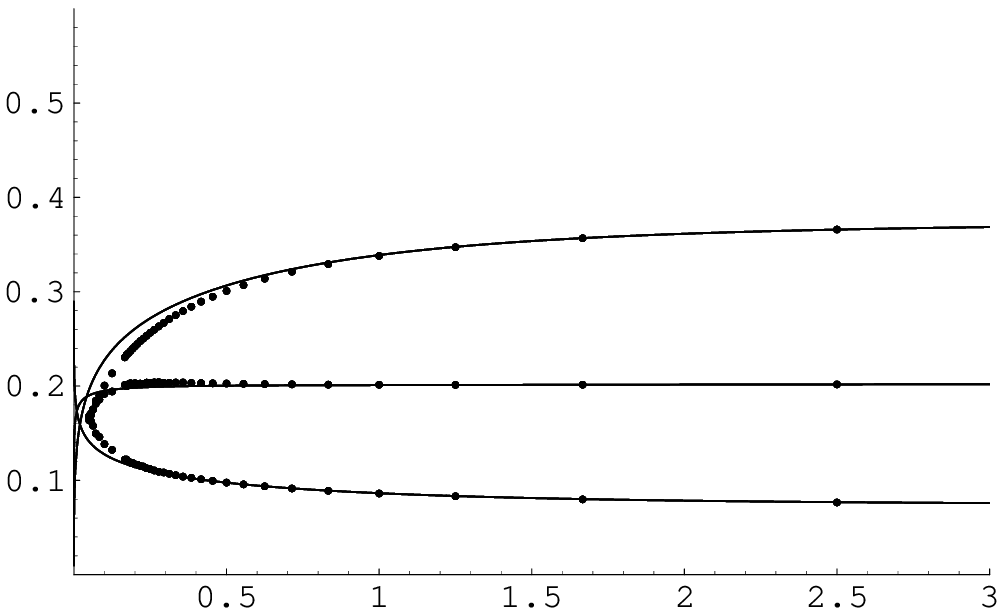}}
\end{picture}
\caption{ {\bf a,b}\quad
The order parameters $C_l$ as function of temperature, for the Random
Charge Model (RCM) with $n_c=3$, $\al=1$, $\eps=\pm1$ resp., and
$P(\la)=\cN(0,1)$. Full lines: annealed theory with fixed mean and
variance. Dots: (quenched) numerical simulations ($N=3000$).
The results are qualitatively correct.}
\lb{fg10}
\end{figure}

\subsection{Annealed Averages for the RIM}

\noi Due to the fact that there are $N(N-1)/2$ disorder parameters
$\la_{ij}$, fixing any overall moment of their probability distribution
does not have any effect. One can e.g. check that the leading order of the
SP values of $\hM$ and $\hS$ is $N(N\m1)/2$, while the
sub-leading order is $\cO(1)$, and the same holds for higher order moments.
This is another indication that the simple annealed average is exact (see
FIG.\ \ref{fg7}).

\beq
\cF_q=\cF_{a_0}\!\!=\ha(\la_0-{\be\la^2\ov2})\De_2\e{\chi\ov\be}\ .
\lb{Fa2i}
\eeq

\noi The divergence of the energy at low temperature is due to the number
of disorder parameters, such that rare, energetically very favorable
combinations of monomers can be formed, to yield arbitrarily low energies.
This would of course not be true for a distribution $P(\la)$ with a support
that is bounded from below.

\section{Conclusions}

\noi We have studied a simple model for random hetero-polymers in
$\infty$-dimensions, the finite dimensional analogues of which had been
studied using various approaches \cite{Ob}-\cite{SG}.
The simplicity and transparency of the model has allowed us to formulate
an exact solution in terms of true order parameters for discrete disorder,
and in terms of self-averaging mesoscopic variables for continuous
disorder. Although the behaviour of finite dimensional models will
obviously be very different, due to the crucial role in the latter of the
chain constraint, our model gives valuable insight into the problems
inherent in the saddle point approximation for the parameters that
occur naturally in a mean field (replica) treatment. Whereas qualitatively
correct for the random hydrophilic-hydrophobic model, the results for the
random charge model are found to be even qualitatively incorrect. For the
random interaction model additional assumptions are needed, but in our
$\infty$-dimensional model at finite temperatures, the simple annealed
average is found to coincide with the quenched result.
Variational approaches have been successful in capturing certain aspects of
the behaviour of random hetero-polymers, but it may well be that the origin
of some of the incongruences between the results obtained by analytical
methods for continuum models and those found numerically for lattice models
\cite{TvMM,GK}, lay in the undue application of the saddle point approximation for
strongly fluctuating quantities. It is not our goal to diminish previous
work which has led to many valuable insights, but to draw the reader's
attention to the inherent risks of the saddle point approximation in random
hetero-polymers.

The Grand Ensemble Approach \cite{Mo,Ku} is a different powerful tool to
describe the case of permuted disorder, but it is technically very hard to
impose independence of the disorder along the sequence. Hence, a good
detailed description of the quenched case based on the Grand Ensemble
Approach is as yet still out of reach in finite dimensional models
\cite{TvMM}.

We have established links between our model, and the Backgammon model
\cite{Ri}, models for polydispersity \cite{poly}, and matching problems
 \cite{VM,Or,MP,N}. The latter in particular suggests that in order to
develop a self-consistent mean field theory for random hetero-polymers, it
may be necessary to find a good set of ``true'' order parameters (see
appendix {\bf B}).

The protein folding problem can be viewed as a geometrically constrained
matching problem for the monomers. Hence, new insights may be gained using
the techniques developed for the matching problem \cite{VM,Or,MP,N},
imposing some weak (geometrical) constraints that can be treated
analytically. The low temperature phase for the random interaction model
certainly needs further inspection. It can be viewed as a generalisation
of the existing results for matching problems, and may be qualitatively
different due to the non-bounded nature of the ground state energy.
Although absent at finite densities, phase transitions may occur when the
density is allowed to become arbitrarily high.

Finally, the study of the dynamics of this model both at finite and at high
density is currently in progress and will be published elsewhere.
\nl
\nl
I am grateful to  A.C.C.~Coolen and R.~K\"uhn for valuable suggestions.


\appendix

\section{Corrections to the Saddle-Point \lb{aA}}

\noi The first correction to the SP for the free energy is easily
calculated, and given by

\beq
-{1\ov2}\log\det(H^*)\simeq{1\ov2\be}\sum^{n_c}_{l=2}C_l\log(l)\ ,
\lb{sadc}
\eeq

\noi where $H^*$ is the Hessian  of $-\be\cF_q$ with respect to the $\n_r$
taken in the SP (\ref{lh}). As we will see now, this correction
is not enough to obtain the correct entropy at high temperature.
\nl
In the limit $\be\to0$ equations (\ref{lh}) simplify to $-i\n_l=\log(l)$,
and the entropy is given by

\beq
\sum_{l=1}^{n_c}f_l\log(l)-\sum_{l=0}^{n_c}C_l\log(C_l)=
-\sum_{l=0}^{n_c}C_l\log({C_l\ov l^l})\ ,
\lb{entw}
\eeq

\noi which is {\em not} correct! From a simple box counting argument one
can show that the non-constant part of the entropy in the high temperature
limit is given by

\beq
-\sum_{l=0}^{n_c}C_l\log(C_l)-\sum_{l=2}^{n_c}C_l\log(l!)=
-\sum_{l=0}^{n_c}C_l\log(l!\ C_l)\ .
\lb{entc}
\eeq

\noi The first term comes from the permutation of the boxes, while the
second one comes from the permutations of monomers within the same box.
To obtain result (\ref{entc}) analytically, one should in principle perform
all the integrations over the $\n_r$ exactly, or equivalently, consider the
whole expansion around the SP. This is an unfeasible task, since
at all temperatures all orders in the expansion around the SP
contribute to the same order in $N$. Instead, at all temperatures, we add
the high temperature correction

\beq
{1\ov\be}\sum^{n_c}_{l=2}C_l\log({l!\ov l^l})
\eeq

\noi to the free energy (\ref{Fqh}) (and the corresponding correction to
(\ref{flh})). In this way, we are at least ensured to obtain the exact
result in the high temperature limit.

\section{Validity of the Annealed Approximation for the RIM. \lb{bB}}

For simplicity and for comparison with the so-called matching problem
\cite{VM,MP,N}, we take $n_c=2$, $\al={1\ov n_c}=\ha$.
We introduce the following order parameters

\bea
d_{a_1,..,a_k}&\ev&{2\ov N}\sum_{i<j=1}^N\lh\prod_{p=1}^k
\de_{r^{a_p}_i,r^{a_p}_j}\rh\ ,\lb{dk}\\
&&1\leq a_1<..<a_k\leq n\quad k=2,..,n,\ ,\nn
\eea

\noi
which have the clear physical meaning of being the fraction of monomers
that are coupled to the same partner in all replicas $a_1,..,a_k$.
The quenched free energy per particle (see (\ref{mZn})) can then be
rewritten as

\bea
\cF_q={\pa\ov\pa n}&&\lv{\la_0\ov2}(\sum_{a=1}^n\De_2^a-n)-{\be\la^2\ov4}
\lh\sum_{a=1}^n\De^a_2+\rp\rp\\
&&\lp\lp\lp +2\sum_{a<b=1}^n(1+d_{ab})\rh-{1\ov\be N}\sum_{m=1}^n
\cS_m\rv\right|_{n=0}\ ,\nn
\eea

\noi
where $\De_a^2=n_c$ (for $\al={1\ov n_c}$).
$\cS_m$ is the entropy of replica $m$ given the configuration of replicas
$1,..,m\m1$ and given the mutual overlaps $d_{a_1,..,a_k}$, $1\!\leq\!
a_1\!<\!..\!<\!a_k\!\leq\! m$\ . To evaluate this entropy we assume
replica symmetry (RS), i.e. $d_{a_1,..,a_k}=d_k,\quad\fa k=1,..,n,$
$\fa a_1\!<\!..\!<\!a_k$. The particles of the $m$-th replica can be
divided into the following non-equivalent groups:

\bea
&&\lh\!\bay{c}m\m1 \\ k\m1\eay\!\rh\hsp{1cm}{\rm groups\ \ of\ \ size:}\nn
\\
&&\hsp{-8mm}Ns_{m,k}\ev N\lv\sum_{l=0}^{m-k}d_{k+l}(-1)^l\lh\!
\bay{c}m\m k\\l\eay\!\rh\rv
~k=1,..,m,
\lb{gro}
\eea

\noi
where $d_1=1$ and $\sum_{k=1}^m\lh\!\bay{c}m\m1\\k\m1\eay\!\rh s_{m,k}=1$.
Note that $k$ indicates the number of replicas (including replica $m$) in
which the couples are equal. Hence, the couples in the groups with $k>1$
are fixed, because they are equal to at least one couple in replicas
$1,..m\m1$. The remaining monomers in $s_{m,1}$ are still free to form
couples with the only restriction that they are not equal to any of the
existing couples in replicas $1,..m\m1$. Hence, the entropy of replica $m$
is composed of both the possible divisions in groups, and the possible
couplings in the last group of size $N s_{m,1}$.
The total number of ways in which $R\gg 1$ particles can be coupled is
denoted by $I_R$, while $I^0_R$ is the number of ways in which $R$
particles can be coupled such that none of the couples is equal to any of
the couples the other replicas. The number of already existing couples
in the other replicas is $c{R\ov2}$ with $1\!\leq\!c\!\leq\!(m\m1)$.
Thus,

\bea
I_R\ev{R!\ov (2!)^{R/2}(R/2)!},~~&~~& S_R\simeq R/2(\log(R)\m1),\nn\\
\lb{cou} \\
\hsp{-5mm}I_R\leq I^0_R\leq I_R\m{cR^2\ov4}I_{R\m2},&&S^0_R\simeq S_R\m
\cO(\log(R)).\nn
\eea

\noi
Using (\ref{gro})-(\ref{cou}), the entropy $\cS_m$ of replica $m$ is given
by

\bea
\cS_m\simeq{N\ov2}&&\lv -\sum_{k=2}^m \lh \bay{c}m\m1\\k\m1\eay\!\rh
s_{m,k}\log(s_{m,k})+\rp \lb{Sm}\\
&&\lp \hsp{-30cm}{1\ov2}\hsp{31cm}+ s_{m,1}(\log(N)-1)\rv,\nn
\eea

\noi
under the assumption that $s_{m,1}=\cO(1)$. The total contribution of $n$
replicas to the entropy $\cS=\sum_{m=1}^n \cS_m$ to leading order in $N$
is now given by

\beq
\cS\simeq {N\ov2}\log(N)\sum_{m=1}^n s_{m,1}=
{nN\ov2}\log(N)\lv1+\sum_{k=2}^\infty{d_k\ov k}\rv,
\eeq

\noi where we have used the equalities

\beq
\sum_{m=0}^p\lh\!\bay{c}m\\k\eay\!\rh=\lh\!\bay{c}p+1\\k+1\eay\!\rh,\quad
\lim_{n\to0}\lh\!\bay{c}n\\k\eay\!\rh\simeq{n(-1)^{k-1}\ov k}\ .
\eeq

\noi
Rescaling the temperature $\be=\be'\sqrt{\log(N)}$, yields the
following state-dependent part of the quenched free energy per particle
to leading order

\beq
\cF_{\rm sd}\simeq {\sqrt{\log(N)}\ov2}\lv{\be'\la^2\ov2}d_2-{1\ov\be'}
\sum_{k=2}^\infty {d_k\ov k}\rv\ .
\lb{fqd}
\eeq

\noi This expression has to be maximized with respect to the $d_k$ with $k$
even, while it has to be minimized with respect to the $d_k$ with $k$ odd.
The $d_k$ cannot be freely extremised as there are constrained from the
fact that all group sizes $s_{m,k}$ are positive: e.g.
$0\!\leq\!d_2\!\leq\!1$, $\max(0,2d_2\m1)\!\leq\!d_3\!\leq\!d_2$, etc.
\nl
For $\be'<{1\ov\la}$ the entropy dominates and the annealed approximation
(i.e. $d_k=0$, $\fa k>1$) is exact. For $\be'\geq{1\ov\la}$ the energy
starts competing and solutions with $d_k>0$ can be found. Eventually, the
system will completely freeze, i.e. $d_k\simeq1$, $\fa k$, but then
expressions (\ref{Sm})-(\ref{fqd}) are no longer valid, since the $s_{m,k}$
are no longer of order $\cO(1)$.
\nl
Although our reasoning remains qualitatively valid for $n_c>2$ and
$\al>{1\ov n_c}$, the details become extremely complicated as the necessary
set of order parameters will be given by

\bea
C_l&\ev&{1\ov N}\sum_{k=1}^N \de_{l,n_{r_i}}\hsp{1cm} l=0,..,n_c \\
d^{\ t}_{a_1,..,a_k}&\ev&{t!\ov N}\sum_{i_1<..<i_t=1}^N\lh\prod_{p=1}^k
\prod_{l=1}^t\de_{r^{a_p}_{i_1},r^{a_p}_{i_l}}\rh, \\
&&\hsp{-1cm} 1\!\leq\!a_1\!<\!..\!<\!a_k\!\leq\!n\ ,
\quad k=2,..,n\ ,\quad t=2,..,n_c\ .\nn
\eea

\noi The latter being the fraction of monomers that are in the same
t-tuple in replicas $a_1,..,a_k$\ .
\end{document}